\documentclass[journal]{IEEEtran}

\IEEEoverridecommandlockouts

\usepackage{cite}
\usepackage{amsmath,amssymb,amsfonts}
\usepackage{algorithm, algorithmic}
\usepackage{graphicx}
\usepackage{textcomp}
\usepackage{xcolor}
\usepackage{comment}
\usepackage{booktabs}
\usepackage{multirow}
\usepackage{tabularx}
\usepackage{makecell}
\usepackage{mathtools}
\usepackage{caption}
\usepackage[normalem]{ulem}
\usepackage{subcaption}
\usepackage{soul}

\def\BibTeX{{\rm B\kern-.05em{\sc i\kern-.025em b}\kern-.08em
    T\kern-.1667em\lower.7ex\hbox{E}\kern-.125emX}}
    
\begin{document}

\title {WiSDoM: Wireless Sparse Decision Transformer with Mixture-of-Experts for Multi-Task Mobile Network Optimization}

\author{\IEEEauthorblockN{Fatih Temiz, Shavbo Salehi, and Melike Erol-Kantarci, \textit{Fellow, IEEE}}

\IEEEauthorblockA{\textit{School of Electrical Engineering and Computer Science, University of Ottawa, Ottawa, Canada} \\
Emails: \{ftemi033, ssale038, melike.erolkantarci\}@uottawa.ca}}

\maketitle

\begin{abstract}
Emerging 6G wireless networks are expected to operate across increasingly diverse deployment scenarios, where variations in network topology, user mobility, traffic demand, and radio conditions challenge the scalability of conventional radio resource management (RRM) methods. While recent offline reinforcement learning (RL) methods have demonstrated strong decision-making capabilities, learning a single policy that performs consistently across heterogeneous wireless environments remains difficult due to conflicting optimization objectives and limited model specialization. These challenges become particularly pronounced in coordinated multipoint (CoMP) transmission, where selecting the optimal serving-cell combination requires sequential decision-making under continuously evolving network conditions.
This paper presents the \emph{Wireless Sparse Decision Transformer with Mixture of Experts} (WiSDoM), a sparse multi-task offline RL framework for adaptive multi-cell selection. Instead of relying on a single shared policy network, WiSDoM combines Decision Transformers (DTs) with a Mixture-of-Experts (MoE) architecture that dynamically activates specialized experts according to the characteristics of each task. This MoE mechanism improves model capacity without proportionally increasing inference cost, mitigates negative transfer across heterogeneous tasks, and enables expert specialization across different tasks.
WiSDoM is trained jointly on diverse network configurations spanning multiple base station and user equipment densities, mobility levels, and scheduler policies. Experimental results show that the proposed framework consistently outperforms heuristic methods, single-task models, and conventional multi-task DTs, improving quality of experience (QoE) by up to 55\% while activating only approximately one-third of the parameters of its dense counterpart during inference. Furthermore, WiSDoM exhibits strong task generalization and efficiently adapts to previously unseen wireless scenarios through few-shot prompting without requiring retraining or fine-tuning.

\end{abstract}


\section{INTRODUCTION}
\IEEEPARstart{A}{rtificial Intelligence (AI)}-native wireless networks are expected to support an unprecedented diversity of devices, services, and performance requirements in the era of hyperconnectivity and pervasive intelligence. Beyond higher data rates, networks evolving toward sixth-generation (6G) systems are anticipated to accommodate massive connectivity, heterogeneous mobility patterns, immersive services and increasingly autonomous network operation. Meeting these requirements calls for an AI-enabled radio access network paradigm capable of continuously adapting its radio resource management (RRM) strategies to rapidly changing channel conditions and service-level objectives \cite{11370176}. Hence, AI-native RRM has emerged as a key component of future wireless systems, offering a data-driven alternative to conventional optimization methods and rigid rule-based control mechanisms \cite{10292755}. Furthermore, AI-driven control is a fundamental enabler of emerging wireless architectures such as the open radio access network (O-RAN), supporting intelligent and adaptive network management for disaggregated networks through the RAN intelligent controllers (RICs) \cite{10329947}.

Among advanced RRM techniques, coordinated multipoint (CoMP) transmission is particularly relevant to dense and heterogeneous deployments \cite{7839266}. CoMP was initially standardized by the 3rd Generation Partnership Project (3GPP) in Release 11 primarily to improve the performance of cell-edge user equipments (UEs). However, its potential extends well beyond its initial interference-management objective. By coordinating multiple base stations (BSs), CoMP can increase spatial diversity, mitigate inter-cell interference, improve coverage, enhance communication reliability through macro-diversity, and improve mobility robustness by reducing handover interruptions across cooperating BSs. These capabilities enable CoMP to support stringent quality of experience (QoE) requirements, making it particularly attractive for ultra-reliable low-latency communications (URLLC), connected vehicles, and industrial automation \cite{9427543,9681835}.
Notably, the performance of CoMP strongly depends on determining which BSs should jointly serve each UE. This multi-cell selection problem becomes increasingly difficult as the network grows because the number of possible serving-cell combinations increases combinatorially with the number of UEs and candidate BSs\cite{schneider2023deepcomp}. Moreover, the optimal association depends on time-varying channel gains, mobility, available radio resources and the scheduling policy employed at each BS.
Exact optimization approaches, such as mixed-integer nonlinear programming (MINLP), can provide optimal or near-optimal solutions but are often difficult to solve directly because of their high computational complexity, motivating the use of decomposition or approximation techniques \cite{9737471}.
Heuristic approaches, including CoMP clustering schemes and signal-to-noise ratio (SNR)-based UE association methods, reduce computational complexity but typically rely on manually designed rules, predefined clustering strategies, or fixed thresholds, limiting their ability to adapt to heterogeneous and time-varying network conditions \cite{article_heuristic, 7839266}. 

Recent advances in reinforcement learning (RL), enabled by deep neural networks in the form of deep reinforcement learning (DRL), have significantly improved the ability to solve complex wireless resource management problems.
DRL algorithms, including proximal policy optimization (PPO), 
\cite{schulman2017ppo}, have been extensively investigated for tasks such as network slicing\cite{10699421}, beam management \cite{10735366}, and particularly for cell-selection in CoMP \cite{schneider2023deepcomp}. Centralized DRL approaches can exploit a global network view to coordinate multi-cell decisions, whereas multi-agent formulations, i.e., multi-agent RL (MARL), can distribute decision-making across BSs or UEs \cite{11432277}. While these approaches outperform conventional heuristics in selected deployment settings, they scale poorly in practice. Also, conventional DRL approaches (e.g., online DRL) require repeated interaction with the environment, which can be costly to implement.
More importantly, conventional DRL policies are generally tied to a predefined state and action space. A change in the number of UEs, the number of BSs, or the underlying network configurations (e.g., scheduler, mobility) may alter the decision problem substantially and require costly retraining \cite{10827032}.

Offline RL offers a promising alternative by learning decision policies from previously collected trajectories without requiring continuous online interaction \cite{xu2022prompt}. In particular, the decision transformer (DT) reformulates RL as a sequence-modeling problem, enabling a policy to predict actions conditioned on historical states, actions and desired returns (i.e., return-to-go (RTG)) \cite{NEURIPS2021_7f489f64}. Moreover, prompt decision transformer (PromptDT) further extends DTs and shows strong performance in multi-task learning and few-shot adaptation by conditioning the policy on short task-specific trajectory demonstrations \cite{xu2022prompt}. This capability is especially attractive for wireless networks, where operating conditions often change frequently after initial deployment. 

Despite these advantages, conventional PromptDT employs a densely shared generative pretrained transformer (GPT) style decoder-only transformer architecture in which the same parameter set is optimized across all tasks. As the diversity of network configurations increases, task-specific gradients can conflict, resulting in negative transfer and degraded multi-task performance. Simply increasing the size of the shared model often provides additional capacity, as in the large language models (LLMs). However, it does not explicitly separate conflicting task-specific knowledge and increases the number of parameters activated during every inference operation \cite{kong2025mastering}. Thus, dense multi-task DTs may exhibit limited task and parameter scalability in highly heterogeneous wireless environments. 

To address these limitations, this paper proposes the \emph{Wireless Sparse Decision Transformer with Mixture of Experts} (WiSDoM) framework for multi-task CoMP multi-cell selection. A lightweight routing mechanism dynamically directs each input trajectory token to a subset of specialized experts, enabling conditional computation that increases model capacity while activating only a fraction of the parameters during inference. The framework is jointly trained on offline trajectories collected from 36 CoMP tasks spanning diverse network configurations, including different numbers of BSs and UEs, UE mobility conditions, and BS scheduling policies (e.g., resource-fair (ReF), proportional-fair (PF)). Consequently, WiSDoM learns a single unified policy across heterogeneous wireless environments instead of requiring independently trained models for each configuration, while trajectory prompts further enable adaptation to previously unseen tasks without retraining or fine-tuning. 

The main contributions of this work are summarized as follows
\begin{itemize}
\item \textit{Multi-Task RL Framework:} We develop a complete multi-task learning framework comprising 36 network configurations with simultaneous variations in UE density, BS deployment, mobility conditions, and scheduling policies.
\item \textit{Computation-Efficient MoE-DT Architecture:} We propose a sparsely activated MoE-DT-based architecture for multi-discrete action spaces, trained using masked cross-entropy loss, auxiliary loss and action masking. The routing mechanism promotes expert specialization across heterogeneous tasks, increases the model capacity, and improves parameter efficiency.
\item \textit{Few-Shot Generalization:} The WiSDoM framework enables adaptation to previously unseen tasks and network conditions through task-specific trajectory prompts, without requiring additional retraining or fine-tuning.
\item \textit{Comprehensive Performance Evaluation:} Extensive experiments show that WiSDoM improves QoE by up to 55\% over heuristic, single-task, and conventional multi-task baselines while using as few as one-third of the active parameters during inference. 
\end{itemize}

\section{Related Work}
\label{sec:related_work}
 \subsection{Generalizable and Multi-Task RL for Wireless Networks}

Early learning-based approaches to address CoMP cell selection relied on online DRL and MARL trained for a fixed network configuration or just extending to UE arrivals \cite{schneider2023deepcomp, 11432277}.
DeepCoMP \cite{schneider2023deepcomp} demonstrated that DRL agents can learn effective multi-cell association policies to maximize user QoE, but its policy must be retrained whenever the number of UEs or BSs changes. Similarly, \cite{11432277} adopts a transformer-based MARL framework (TransfQMix) for CoMP cell selection, improving inter-agent coordination and scalability to varying UE populations. Nevertheless, it still relies on online MARL training and does not learn a unified offline policy across heterogeneous tasks. Beyond transformer-based RL architectures, recent studies have also begun leveraging generative AI (GenAI) to improve wireless network optimization. The LLM-Curriculum approach~\cite{10682015} improves the training of PPO-based CoMP association by using an LLM (e.g., GPT-4) to construct a curriculum over increasingly complex UE configurations. Although curriculum learning can improve training efficiency, the resulting policy remains based on online environment interaction. 

Several studies have instead formulated wireless resource management as offline sequence modeling using DTs rather than DRL.
In \cite{ali2024explainability}, the authors explore explainability for DT-based CoMP cell selection using both intrinsic and post-hoc explainable AI (XAI) methods. While the study provides valuable insights into the DT's decision-making process, it considers only a single network configuration and does not address multi-task learning or generalization across heterogeneous environments. On the other hand, ~\cite{10839243} uses DT-based control for joint intelligent reflecting surface (IRS) phase-shift and unmanned aerial vehicle (UAV) trajectory optimization, demonstrating multi-objective control within a wireless resource-management problem under predefined system configurations.
Also, the authors in ~\cite{11080254} combine PromptDT with federated learning to maximize the QoE of customized mobile edge computing (MEC)-based virtual-reality services across heterogeneous user environments. 

More closely related to task generalization, the methods in~\cite{10827032} and~\cite{11021485} use PromptDT to adapt across varying numbers of BSs and UEs, respectively. The former targets BS energy management, whereas the latter enhances PromptDT with an attention mechanism to overcome limitations due to zero-padding and jointly optimizes UAV trajectory planning and user scheduling in UAV-assisted Internet-of-Things (IoT) networks.
Authors in ~\cite{Temiz2026EdgeLearning} proposed federated split decision transformer (FSDT), which distributes DT training between the cloud and MEC servers and uses MEC-specific modules to support environment customization while reducing computational burden on MECs.
Similarly, authors in ~\cite{Temiz2026Generalizable} proposed a PromptDT-based framework that jointly generalizes across variations in UE count, BS count, and scheduling policy in CoMP; however, it retains a densely shared transformer backbone on a limited number of tasks and therefore does not explicitly address parameter-efficient scaling or expert specialization across tasks. 
 
Table~\ref{tab:related_work} summarizes the main differences between these studies and WiSDoM. It analyzes all related works from online versus offline RL paradigms, proposed algorithms, wireless network domain, objective of the approach and whether MoE is used or not.

\begin{table*}[t]
\centering
\caption{Literature review of generalizable, multi-task RL, DT-based and MoE-based frameworks for wireless networks.}
\label{tab:related_work}
\renewcommand{\arraystretch}{1.3}
\setlength{\tabcolsep}{3.8pt}
\footnotesize

\begin{tabularx}{\textwidth}{|l|X|c|c|c|c|c|}
\hline
\textbf{Reference} &
\makecell{\textbf{Method} \textbf{Summary}} &
\makecell{\textbf{Online }\textbf{RL}} &
\makecell{\textbf{Off.} \textbf{RL}} &
\makecell{\textbf{Multi-}\textbf{Task}} &
\makecell{\textbf{Task} \textbf{Variation}} &
\textbf{MoE} \\
\hline

\cite{schneider2023deepcomp} & DRL for adaptive CoMP cell selection & \checkmark & -- & -- & -- & -- \\
\hline
\cite{11432277} & Transformer-based MARL approach for cooperative CoMP  & \checkmark & -- & -- & -- & -- \\
\hline
\cite{10682015} & GPT-4-assisted PPO for multi-task CoMP & \checkmark & -- & -- & UE Count & -- \\
\hline
\cite{ali2024explainability} & XAI-assisted DT for CoMP & -- & \checkmark & -- & -- & -- \\
\hline
\cite{10839243} & DT for joint IRS-UAV optimization & -- & \checkmark & \checkmark & IRS+UAV & -- \\
\hline
\cite{11112781} & Hierarchical DT for AI-enabled RAN intent management & -- & \checkmark & -- & -- & -- \\
\hline
\cite{11080254} & Federated PromptDT for personalized MEC-VR optimization & -- & \checkmark & \checkmark & User Env. & -- \\
\hline
\cite{10827032} & PromptDT for generalizable BS energy optimization & -- & \checkmark & \checkmark & BS Count & -- \\
\hline
\cite{11021485} & Attention enhanced PromptDT for UAV-IoT control & -- & \checkmark & -- & UE Count & -- \\
\hline
\cite{app16041823} & MoE-enhanced DRL for adaptive RAN slicing & \checkmark & -- & \checkmark & SLA weights & \checkmark \\
\hline
\cite{11240212} & MoE-based meta-RL for adaptive MAC protocols & \checkmark & -- & \checkmark & MAC protocols & \checkmark \\
\hline
\cite{10592370} & LLM-gated MoE for network utility optimization & \checkmark & -- & \checkmark & QoS intent & \checkmark \\
\hline
\cite{10901084} & Distributed MoE inference for wireless LLMs & -- & -- & -- & Expert placement & \checkmark \\
\hline
\cite{11303878} & MoE-based distributed GenAI for mobile edge Metaverse
& -- & -- & \checkmark & GenAI tasks & \checkmark \\
\hline
\cite{11417148} & Personalized federated MoE for heterogeneous AMC & -- & -- & \checkmark & Data heterogeneity & \checkmark \\
\hline
\cite{11527015} & Hybrid expert quantization for edge MoE deployment & \checkmark & -- & -- & Request distribution & \checkmark \\
\hline
\cite{Temiz2026EdgeLearning} & Federated split DT for MEC-VR & -- & \checkmark & \checkmark & Network characteristics & -- \\
\hline
\cite{Temiz2026Generalizable} & PromptDT for generalizable CoMP & -- & \checkmark & \checkmark & BS, UE, Scheduler & -- \\
\hline
\makecell{\textbf{WiSDoM}\\\textbf{(ours)}} &
\textbf{Sparse MoE-enhanced PromptDT for generalizable CoMP} &
-- & \checkmark & \checkmark &
\textbf{BS, UE, Mobility, Scheduler} &
\checkmark \\
\hline

\end{tabularx}
\end{table*}

\subsection{MoE for Scalable Sequence Modeling}
Scaling dense transformers uniformly, i.e., activating all parameters for every input, becomes computationally prohibitive as model capacity grows. However, expanding capacity remains attractive because larger sequence models have 
been shown to implicitly perform many tasks without task-specific supervision, first demonstrated at scale by GPT-2 \cite{radford2019language}.
MoE architectures provide a scalable form of conditional computation for transformer-based models. Rather than activating the same parameters for every input token, a learned router selects a small subset of specialized expert networks. This design increases total model capacity while keeping the number of active parameters per input comparatively small~\cite{fedus}. Expert specialization can also reduce interference (i.e., gradient conflict) among heterogeneous data distributions by sequential training of each expert.
This architecture has been adopted by several high-capacity frontier language models. 
DeepSeek-V3 contains 671 billion parameters while activating only 37 billion per token through DeepSeekMoE \cite{deepseekai2025deepseekv3technicalreport}. 
NVIDIA's Nemotron 3 Super and Ultra further combine sparse MoE routing with hybrid mamba-attention backbones, activating 12 billion of 120 billion and 55 billion of 550 billion parameters, respectively. These models demonstrate that conditional expert activation can substantially increase representational capacity while maintaining computationally efficient inference \cite{nvidia2026nemotron3super}.

However, these models employ MoE primarily for language modeling, where experts process token representations to predict subsequent tokens. It is important to note that experts in these LLMs are not generally assigned explicit high-level functions such as translation, summarization, or coding. Instead, routing is learned at the token level, and expert specialization emerges from the training data and optimization process. 
MoE has also been investigated for scalable multi-task control with RL. Authors in \cite{kong2025mastering} augment a PromptDT backbone with grouped experts and a multi-stage training procedure to improve performance as the number of control tasks increases.
Additionally, \cite{wu2025mixtureofexperts} introduces token- and task-wise MoE for in-context RL, which incorporates sparse MoE routing to improve multi-task decision making across heterogeneous control tasks.
Their evaluation, however, focuses on standard RL benchmarks and does not consider wireless-RRM characteristics such as time-varying channels, mobility, heterogeneous schedulers, constrained multi-discrete actions, or network-level QoE objectives.

\subsection{MoE for Wireless Networks}
As wireless networks become increasingly heterogeneous and AI-native, single dense models face growing challenges in scaling across diverse optimization tasks. The MoE architecture addresses these limitations through conditional computation, which enables larger model capacity without proportionally increasing inference cost, making MoE particularly attractive for resource-constrained wireless systems. Authors in ~\cite{xu2026moesurvey} provide a comprehensive review of MoE architectures for wireless communications, examining how expert specialization addresses core network challenges.
\begin{itemize}
\item Increased model capacity through experts tailored to distinct tasks (e.g., adaptive modulation classification (AMC), interference cancellation, channel estimation).
\item Computational efficiency via sparse expert activation suited to resource-constrained edge/IoT deployments.
\item Dynamic feature learning for time-varying conditions such as spectrum allocation and beamforming. 
\end{itemize}
Authors in \cite{xu2026moesurvey} further discuss MoE's integration with GenAI and RL across channel prediction, resource allocation, and network security, along with multi-task and multimodal specialization (e.g., SNR-regime-specific experts for AMC, modality-specific experts for radar/LiDAR/visual data). 
In \cite{11417148}, a federated MoE combines shared and client-specific experts through dynamic gating to address data heterogeneity in AMC, while \cite{11527015} optimizes expert quantization (i.e., expert-wise mixed-bit quantization) and request assignment for efficient MoE deployment at resource-constrained edge servers. Similarly, \cite{11303878} distributes GenAI tasks among specialized edge experts, demonstrating the potential of MoE-based task decomposition and expert specialization for reducing computational and serving costs.

In \cite{app16041823}, the authors pre-train and freeze multiple expert policies under diverse slice service level agreement (SLA) weight preferences (e.g., VoLTE, URLLC, etc.), then train a step-level DRL-based gating network to fuse expert actions for fast adaptation to unseen SLA configurations in RAN slicing. While it demonstrates that MoE gating generalizes well to unseen SLA weight vectors, its experts are trained independently offline and fused only at the gating level during online interaction rather than being jointly optimized within a single sequence model.
MoE is also incorporated into a context-based meta-RL framework for media-access control (MAC) in heterogeneous networks \cite{11240212}. The expert modules improve latent-context representation and support rapid adaptation to previously unseen MAC-protocol environments. Unlike \cite{app16041823}, the experts and gating mechanism are jointly trained through meta-RL; nevertheless, MoE is used primarily for context representation within a step-wise policy rather than for autoregressive modeling of extended decision trajectories. 

Authors in ~\cite{10592370} replace a conventional learned gate with an LLM. Given a textual description of a user's QoS requirements, the LLM selects and weights independently pre-trained DRL experts, such as outage probability minimization and data rate maximization policies, to optimize network-provider utility. Compared to our approach, this work uses MoE externally as an orchestration mechanism over multiple independent RL models, while WiSDoM uses MoE internally within a single model.
This approach also incurs LLM inference overhead and relies on natural-language task specifications.
Taken together, \cite{app16041823,11240212,10592370} demonstrate that MoE is
an effective mechanism for adapting to task heterogeneity in wireless RL.
However, these approaches do not combine sparse expert routing with long-horizon offline trajectory modeling over a large and structurally diverse set of wireless control tasks. 

\textit{Uniqueness of WiSDoM framework:} To the best of our knowledge, WiSDoM is the first framework that combines sparse MoE routing with PromptDT-based offline RL for the wireless communications domain (i.e., particularly CoMP). It jointly learns across a large number of diverse network configurations and supports prompt-based adaptation to unseen tasks, and shows strong few-shot prediction performance.

\section{SYSTEM MODEL AND PROBLEM FORMULATION}
\label{sec:system_model}

\subsection{System Model} 

In this paper, we consider a downlink (DL) CoMP transmission scenario with joint transmission and coordinated scheduling. The wireless network consists of $M$ UEs and $N$ BSs, where both $M$ and $N$ can be adjusted as illustrated in Fig. \ref{fig:systemmodel}. The system is assumed to operate over discrete time steps $t = 1, 2, \dots, T$ within a bounded two-dimensional area where possible configurations are listed in Table \ref{tab:task_config}. All BSs are assumed to transmit synchronously.
UE mobility follows the random waypoint model, where each UE moves with a constant velocity and periodically selects random destinations within the environment. Wireless channel propagation is modeled using the Okumura-Hata path loss model \cite{851585}. 
 Orthogonal resource allocation is assumed within each BS, such that different UEs are assigned different resource blocks (RBs), thereby eliminating intra-cell interference. Furthermore, neighboring BSs are assumed to operate on orthogonal or slightly shifted frequency resources, and therefore inter-cell interference is neglected \cite{schneider2023deepcomp}. At each time step, the SNR between UE $u_j$ and BS $b_i$ is denoted by $\rho_{ij}(t)$. A UE can establish or maintain a DL connection with a BS only if the measured SNR exceeds a predefined threshold $\rho_{\min}$.
The path loss between BS $b_i$ and UE $u_j$ at distance $d_{ij}(t)$ (km) is modeled using the Okumura-Hata urban propagation model:

\begin{equation}
    \begin{aligned}
        \mathrm{PL}_{ij}(t) = \, & 69.55 - a(h_{u_j}) + 26.16\log_{10}f_c - 13.82\log_{10}h_{b_i} \\
        & + \left(44.9 - 6.55\log_{10}h_{b_i}\right)\log_{10}d_{ij}(t)
    \end{aligned}
\end{equation}
\begin{equation}
    a(h_{u_j}) = 0.8 + (1.1\log_{10}f_c - 0.7)h_{u_j} - 1.56\log_{10}f_c
\end{equation}
where $f_c$ is the carrier frequency (MHz), $h_{b_i}$ and $h_{u_j}$ denote the BS and UE antenna heights (m), and $a(h_{u_j})$ is the mobile antenna height correction factor for urban environments. The received SNR $\rho_{ij}(t)$ is then computed from $\mathrm{PL}_{ij}(t)$ together with the transmit power and noise floor. 

Multiple BSs are allowed to simultaneously serve a single UE, enabling CoMP transmission. BS resources, i.e., physical resource blocks (PRBs), are allocated among their connected UEs using either a ReF or PF scheduling. Under ReF scheduling, each BS allocates its available PRBs equally among its connected UEs. Under PF scheduling, PRBs are distributed by balancing UEs’ instantaneous data rates with their historical throughput to achieve long-term fairness \cite{schneider2023deepcomp}.
The achievable Shannon DL data rate between BS $b_i$ and UE $u_j$ at time $t$ is denoted by $D_{ij}(t)$, and depends on the channel conditions and scheduling policy. The total DL data rate achieved by UE $u_j$ is then:
\begin{equation}
D_j(t)=\sum_{b_i \in \mathcal{C}_j(t)} D_{ij}(t),
\end{equation}
where $\mathcal{C}_j(t)$ denotes the set of BSs serving UE $u_j$ at time $t$.

To quantify the QoE of the UEs, we employ a bounded logarithmic utility function that maps the achieved DL data rate to a finite utility range. Let $D_j(t)$ denote the aggregated DL data rate received by UE $u_j$ at time $t$. The instantaneous utility of UE $u_j$ is defined as :
\begin{equation}
U_j(t)=\min\!\left\{U_{\max},\;\max\!\left\{U_{\min},
\;
w_1 \frac{\log\!\left(w_2 + D_j(t)\right)}{\log(w_3)}
\right\}
\right\},
\end{equation}
where $w_1$, $w_2$, and $w_3$ are configurable scaling coefficients, and $U_{\min}$ and $U_{\max}$ denote the lower and upper bounds of the utility function, respectively \cite{schneider2023deepcomp}. The clipping operation ensures that the utility remains within a bounded interval, and the rationale of this QoE formulation is to capture the diminishing returns of user-perceived satisfaction with increasing data rates.

The objective of the UE association problem is to determine the UE-BS connectivity decisions over time to maximize the long-term average QoE across all active UEs, subject to connectivity and SNR constraints, and it can be expressed as follows:
\begin{subequations}\label{eq:qoe_optimization}
\begin{equation}\label{eq:qoe_optimization_obj}
\max_{\{x_{ij}(t)\}} \;
\lim_{T\to\infty}\frac{1}{T}\sum_{t=1}^{T}\frac{1}{M(t)}
\sum_{j\in\mathcal{U}(t)} U_j(t),
\tag{\theparentequation}
\end{equation}
\begin{equation}\label{eq:qoe_optimization_bin}
x_{ij}(t)\in\{0,1\},
\quad \forall j\in\mathcal{U}(t),\ \forall i\in\mathcal{B},\ \forall t
\end{equation}
\begin{equation}\label{eq:qoe_optimization_snr}
x_{ij}(t)=0 \ \text{if}\ \rho_{ij}(t)\le \rho_{\min},
\quad \forall j\in\mathcal{U}(t),\ \forall i\in\mathcal{B},\ \forall t
\end{equation}
\begin{equation}\label{eq:qoe_optimization_conn_limit}
\sum_{i\in\mathcal{B}} x_{ij}(t) \le N,
\quad \forall j\in\mathcal{U}(t),\ \forall t
\end{equation}
\end{subequations}
where Eq. \eqref{eq:qoe_optimization_obj} maximizes the long-term average QoE of all active UEs. Eq. \ref{eq:qoe_optimization_bin} defines the binary UE-BS association variables, and Eq. \ref{eq:qoe_optimization_snr} enforces that a UE–BS connection can only be established when the SNR constraint is satisfied.
Also, Eq. \ref{eq:qoe_optimization_conn_limit} ensures that the total number of simultaneous connections for each UE does not exceed the number of BSs. 

\subsection{Problem Formulation}

Due to the stochastic evolution of user mobility and channel conditions, we model the multi-cell selection problem as a Markov Decision Process (MDP), where future system states depend only on the current state and action.
The RL agent's MDP is defined as follows:
\paragraph{State Space}
At time step $t$, the system state captures the DL connectivity, channel quality, and service utility of all active UEs. For a system with $M(t)$ active UEs and $N$ BSs, the state is defined as:
\begin{equation}
\mathbf{s}_t =
\Big[
\mathbf{x}_j(t),\;
\boldsymbol{\rho}_j(t),\;
\hat{U}_j(t)
\Big]_{j\in\mathcal{U}(t)},
\end{equation}
$\mathbf{x}_j(t)\in\{0,1\}^{N}$ denotes the binary connection vector where the $n$-th element equals to $1$ if UE $u_j$ is connected to BS $b_n$, and $0$ when it is not connected. $\boldsymbol{\rho}_j(t)\in\mathbb{R}^{N}$ represents the normalized SNRs between UE $u_j$ and all BSs, and $\hat{U}_j(t)\in[-1,1]$ is the scaled utility of UE $u_j$.

\paragraph{Action Space}
To limit protocol overhead and signaling complexity, each UE is allowed to modify at most one connection per time step. Accordingly, the action for UE $u_j$ at time $t$ is defined as:

\begin{equation}
a_j(t)\in\{0,1,\dots,N\},
\end{equation}
where $a_j(t)=i\in\{1,\dots,N\}$ indicates that UE $u_j$ toggles its connection status with BS $b_i$, i.e., a connection is established if none exists or released otherwise, provided that the SNR constraint is satisfied. Alternatively, $a_j(t)=0$ denotes a non-operation, where all existing connections of UE $u_j$ remain unchanged. Then the joint action at time $t$ is given by $\mathbf{a}_j = \{a_j(t)\}_{j\in\mathcal{U}(t)}$.

\paragraph{Reward Function}
The reward function at the time step $t$ is defined as the average utility of all active UEs, reflecting the overall system's QoE:
\begin{equation}\label{eq:rewarddd}
r_t =
\frac{1}{M(t)}\sum_{j\in\mathcal{U}(t)} \hat{U}_j(t),
\end{equation}
where the reward formulation in Eq. \ref{eq:rewarddd} encourages policies that maximize the long-term average QoE across users.

\section{WISDOM for Multi-Task Learning in CoMP}
\label{sec:wisdom}
\subsubsection{Multi-Task RL}
RL is conventionally formulated as an MDP $(\mathcal{S},\mathcal{A},\mathcal{P},\mathcal{R},\gamma,d_0)$, with state space $\mathcal{S}$, action space $\mathcal{A}$, transition dynamics $\mathcal{P}(s'|s,a)$, reward function $\mathcal{R}$, discount factor $\gamma$, and initial state distribution $d_0$. In multi-task RL (MTRL), a family of tasks $\mathcal{T}_i \sim p(\mathcal{T})$ is considered, where each task induces a distinct MDP $(\mathcal{S}^{\mathcal{T}_i},\mathcal{A}^{\mathcal{T}_i},\mathcal{P}^{\mathcal{T}_i},\mathcal{R}^{\mathcal{T}_i},\gamma,d_0^{\mathcal{T}_i})$ with potentially different state/action dimensionality, dynamics, and reward structure. Rather than solving a single MDP, the objective is to learn a shared policy that generalizes across the task distribution:
\begin{equation}
    \pi^{\star} = \arg\max_{\pi}\ \mathbb{E}_{\mathcal{T}_i \sim p(\mathcal{T})}\ \mathbb{E}_{a_t\sim \pi}\Big[\sum_{t=0}^{\infty}\gamma^t r_t^{\mathcal{T}_i}\Big].
\end{equation}
In the offline setting, a static multi-task dataset $\mathcal{D} = \bigcup_{i} \mathcal{D}_i$ is provided, partitioned into per-task subsets $\mathcal{D}_i$ collected by a behavior policy, and training tasks $\mathcal{T}^{\mathrm{train}}$ are used to learn a policy that is then evaluated on both seen and unseen tasks $\mathcal{T}^{\mathrm{eval}}$, without further environment interaction. In wireless CoMP multi-cell selection, task heterogeneity naturally arises from varying numbers of BSs and UEs, UE mobilities, scheduler policies (e.g., ReF/PF), and network topologies, each inducing a distinct state/action space and reward landscape that a shared policy must generalize across.

\subsubsection{DT and PromptDT}
Offline RL seeks to learn a policy entirely from a fixed dataset $\mathcal{D}=\{(s,a,s',r)\}$ without online interaction, which is well suited to wireless settings where exploratory interaction with live infrastructure is costly \cite{10839243}. DT reframes offline RL as a sequence modeling problem, replacing explicit value-function estimation or policy optimization with autoregressive prediction over trajectories of RTG, state, and action tokens:
\begin{equation}
    \tau = \left(\hat{R}_0, s_0, a_0, \hat{R}_1, s_1, a_1, \ldots, \hat{R}_T, s_T, a_T\right),
\end{equation}
where $\hat{R}_t = \sum_{t'=t}^{T} r_{t'}$ is the undiscounted RTG. Tokens are embedded into a shared latent space, augmented with positional embeddings, and processed by masked causal self-attention blocks following a GPT-style decoder, so that action prediction is cast as conditional next-token generation given a desired target return\cite{NEURIPS2021_7f489f64}.

While DT generalizes poorly across tasks with a single shared parameterization, PromptDT extends the architecture with short task-identifying trajectory prompts $\tau_i^{\star} = (\hat{r}^{\star}, s^{\star}, a^{\star})$ of length $K^{\star}$, sampled from near-optimal demonstrations $\mathcal{P}_i$, and prepended to a $K$-step history segment ($K^{\star} < K$) to form the input $\tau_i^{\mathrm{input}} = (\tau_i^{\star}, \tau_i)$. The prompt encodes just enough information to identify the task, and the model learns to infer task identity implicitly, enabling few-shot generalization to unseen tasks without additional retraining, analogous to in-context learning (ICL) in LLMs\cite{xu2022prompt, NEURIPS2020_1457c0d6}. ICL with LLMs has shown strong promise in several wireless network problems such as network intrusion detection \cite{10901312}.

Despite these benefits, a single shared backbone still suffers from gradient conflicts as the number and heterogeneity of tasks grows, since all tasks compete for the same parameters \cite{kong2025mastering}. This motivates augmenting the PromptDT backbone with an MoE architecture, where dedicated expert modules absorb task-specific specialization while the backbone retains shared cross-task knowledge, alleviating gradient interference across heterogeneous CoMP configurations. This forms the basis of the proposed WiSDoM framework, introduced next.

\begin{algorithm}[ht]
\footnotesize
\caption{Backbone and Expert Training (Stages 2--3)}
\label{alg:wisdom-stage12}
\begin{algorithmic}[1]

\REQUIRE Training tasks $\mathcal{T}^{\mathrm{train}}$, task groups
$\{\mathcal{G}_1,\ldots,\mathcal{G}_C\}$, PromptDT backbone $f_{\theta}$,
offline datasets $\{\mathcal{D}_i\}$, demonstrations $\{\mathcal{P}_i\}$,
training iterations $N_1,N_2$, per-task batch size $M$,
number of transformer blocks $L$

\STATE \textbf{Stage 2: Multi-Task Backbone Training}
\FOR{$n = 1$ to $N_1$}
    \FOR{each task $\mathcal{T}_i \in \mathcal{T}^{\mathrm{train}}$}
        \FOR{$m = 1$ to $M$}
            \STATE Sample trajectory $\tau_{i,m} \sim \mathcal{D}_i$ and prompt
            $\tau_{i,m}^{\star} = \text{GetPrompt}(\mathcal{T}_i,\mathcal{P}_i)$
            \STATE Construct input
            $\tau_{i,m}^{\mathrm{input}} = (\tau_{i,m}^{\star},\tau_{i,m})$
        \ENDFOR
        \STATE Form task minibatch $\mathcal{B}_i^M = \{\tau_{i,m}^{\mathrm{input}}\}_{m=1}^{M}$
    \ENDFOR
    \STATE Form multi-task batch $\mathcal{B} = \bigcup_{\mathcal{T}_i} \mathcal{B}_i^M$
    \STATE Predict actions $a^{\mathrm{pred}} = f_{\theta}(\tau^{\mathrm{input}})$, $\forall \tau^{\mathrm{input}} \in \mathcal{B}$
    \STATE Compute masked cross-entropy loss $\mathcal{L}_{\mathrm{CE}}$
    \STATE Update backbone: $\theta \leftarrow \theta-\alpha_1\nabla_{\theta}\mathcal{L}_{\mathrm{CE}}$
\ENDFOR

\STATE \textbf{Stage 3: Group-Specific Expert Training}
\STATE Freeze the shared backbone parameters $\theta$
\FOR{each task group $\mathcal{G}_c$, $c=1,\ldots,C$}
    \STATE Initialize one expert $e_{\phi_c}^{(\ell)}$ in each transformer block $\ell=1,\ldots,L$
    \FOR{$n = 1$ to $N_2$}
        \STATE Sample a batch $\mathcal{B}_c$ from tasks in $\mathcal{G}_c$
        \FOR{each block $\ell=1,\ldots,L$}
            \STATE Combine with FFN in parallel:
            $x_t^{(\ell)} \leftarrow x_t^{(\ell)} + h_{t,\mathrm{FFN}}^{(\ell)} + e_{\phi_c}^{(\ell)}(x_t^{(\ell)})$
        \ENDFOR
        \STATE Predict actions $a^{\mathrm{pred}} = f_{\theta,\phi_c}(\mathcal{B}_c)$
        \STATE Compute masked cross-entropy loss $\mathcal{L}_{\mathrm{CE}}^{(c)}$
        \STATE Update only expert parameters: $\phi_c \leftarrow \phi_c - \alpha_2 \nabla_{\phi_c}\mathcal{L}_{\mathrm{CE}}^{(c)}$
    \ENDFOR
\ENDFOR

\STATE \textbf{return} backbone $\theta$, experts $\Phi=\{\phi_1,\ldots,\phi_C\}$

\end{algorithmic}
\end{algorithm}
\begin{figure*}[h!]
    \centering
    \includegraphics[width=2\columnwidth]{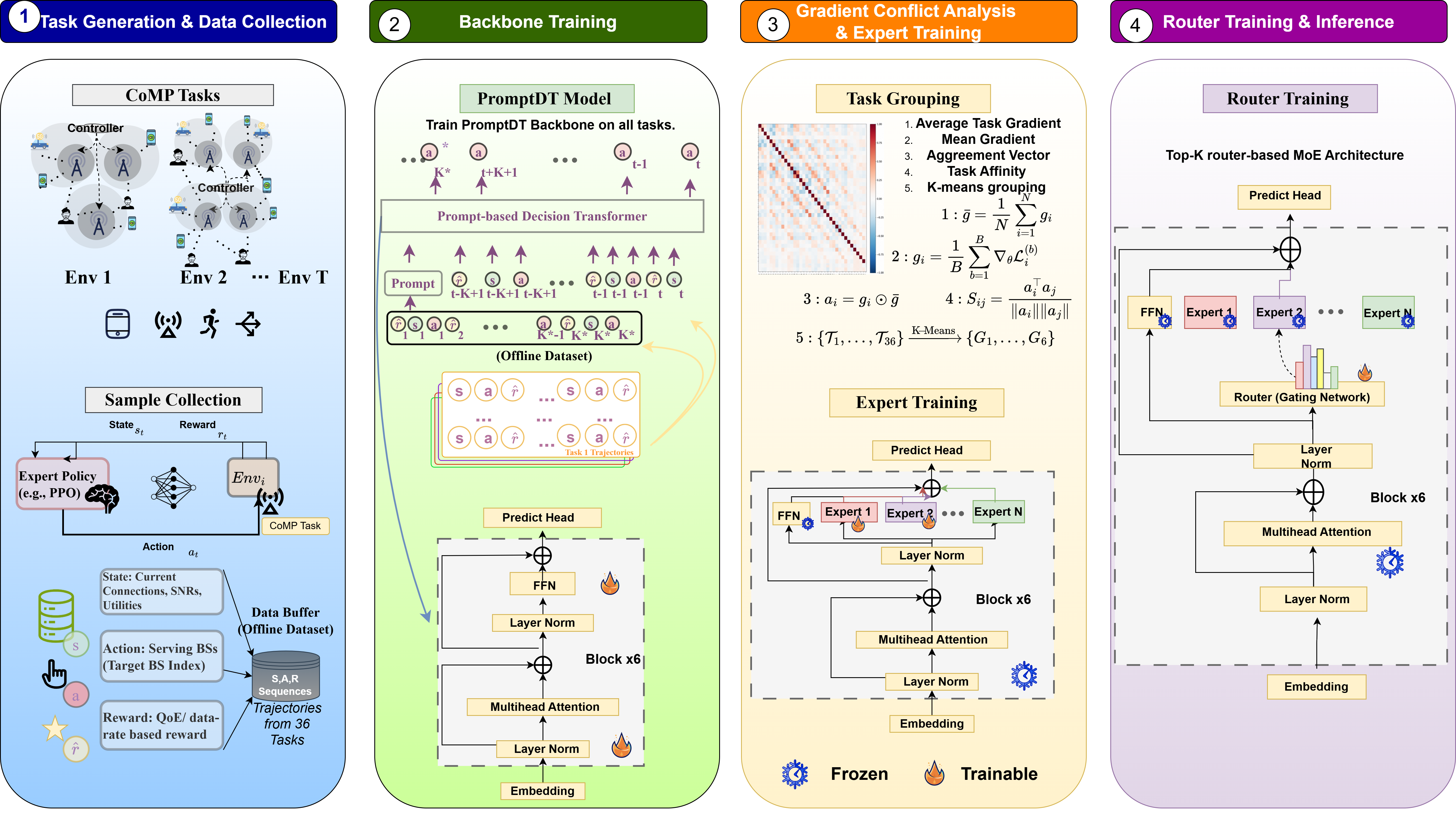}
    \caption{Overview of WiSDoM Framework}
    \label{fig:systemmodel}
\end{figure*} 
\subsection{WiSDoM Framework}
The WiSDoM framework is illustrated in Fig. \ref{fig:systemmodel} including four main steps. Also, Algs. \ref{alg:wisdom-stage12} and \ref{alg:wisdom-stage3} break down the framework in detail. The first part of the Fig. \ref{fig:systemmodel} is the offline setting where the global data buffer consists of RL trajectories formed over diverse tasks. Later, in backbone training, the PromptDT model is trained across all tasks utilizing historical trajectories with prompts. Third, the trained backbone is used to analyze inter-task gradient conflicts, from which agreement vectors are computed to group similar tasks. An expert is then assigned to each task group by inserting task-specific feedforward networks (FFN) into the transformer blocks, and each expert is fine-tuned using only the data from its corresponding task group. 
Finally, the backbone and experts are frozen, and only the router (i.e., gating network) is optimized to enable dynamic token-level expert selection. During inference, each transformer block performs self-attention followed by sparse expert routing, where the router selects the top-ranked expert(s) for each token. The resulting representations are then passed through the prediction heads to generate the final outputs.

\begin{algorithm}[h]
\footnotesize
\caption{Router Training (Stage 4) and Evaluation}
\label{alg:wisdom-stage3}
\begin{algorithmic}[1]
\REQUIRE Trained backbone $f_{\theta}$, trained experts $\Phi=\{\phi_1,\ldots,\phi_C\}$, evaluation tasks $\mathcal{T}^{\mathrm{eval}}$, training iterations $N_3$, routing sparsity $K_{\mathrm r}$, target returns $G^\star$, episode length $T$, prompt length $K^\star$, auxiliary-loss coefficient $\lambda_{\mathrm{aux}}$
\STATE \textbf{Router Training}
\STATE Freeze backbone $\theta$ and experts $\Phi$; initialize router $r_{\psi}^{(\ell)}$ at each block $\ell$
\FOR{$n=1$ to $N_3$}
\STATE Sample a multi-task batch $\mathcal{B}$
\FOR{each block $\ell=1,\ldots,L$}
\STATE Given normalized token $x_t^{(\ell)}$, compute $z_t^{(\ell)}=r_{\psi}^{(\ell)}(x_t^{(\ell)})$ and $p_t^{(\ell)}=\operatorname{Softmax}(z_t^{(\ell)})$
\STATE Select $\mathcal{E}_t^{(\ell)} =\operatorname{TopK}(p_t^{(\ell)},K_{\mathrm r})$ and construct normalized sparse weights $\hat{w}_t^{(\ell)}$
\STATE Apply STE: $w_t^{(\ell)} = p_t^{(\ell)} + \operatorname{sg}\left(\hat{w}_t^{(\ell)}-p_t^{(\ell)}\right)$
\STATE Compute MoE output: $h_{t,\mathrm{MoE}}^{(\ell)} = \sum_{e\in\mathcal{E}_t^{(\ell)}} w_{t,e}^{(\ell)} e_{\phi_e}^{(\ell)}(x_t^{(\ell)})$
\STATE Combine with the parallel FFN path: $x_t^{(\ell)} \leftarrow x_t^{(\ell)} + h_{t,\mathrm{FFN}}^{(\ell)} + h_{t,\mathrm{MoE}}^{(\ell)}$
\ENDFOR
\STATE Predict actions $a^{\mathrm{pred}}=f_{\theta,\Phi,\psi}(\mathcal{B})$
\STATE Compute $\mathcal{L}_{\mathrm{router}} = \mathcal{L}_{\mathrm{CE}}^{\mathrm{router}} + \lambda_{\mathrm{aux}}\mathcal{L}_{\mathrm{aux}}$
\STATE Update only the router: $\psi \leftarrow \psi - \alpha_3\nabla_{\psi} \mathcal{L}_{\mathrm{router}}$
\ENDFOR
\STATE \textbf{Evaluation}
\FOR{each task $\mathcal{T}_j\in\mathcal{T}^{\mathrm{eval}}$}
\STATE Initialize trajectory history $\tau$, target return $g=G_j^\star$, and prompt $\tau_j^\star = \operatorname{GetBestPrompt}(\mathcal{T}_j,\mathcal{P}_j)$ of length $K^\star$ 
\FOR{$t=1$ to $T$}
\STATE $a_t = f_{\theta,\Phi,\psi}\left((\tau_j^\star,\tau)\right)[-1]$
\STATE Execute $a_t$, observe $s_{t+1}$ and $r_t$; update $g\leftarrow g-r_t$
\STATE Append $[s_{t+1},a_t,g]$ to $\tau$
\ENDFOR
\ENDFOR
\end{algorithmic}
\end{algorithm}

\subsubsection{Trajectory Collection}
DT-based models are trained on offline trajectories consisting of state, action, and RTG sequences. These trajectories can be collected using any behaviour policy, including RL agents, heuristic algorithms, or human demonstrations \cite{xu2022prompt}. In this work, the offline dataset is generated using expert policies trained with the PPO algorithm in diverse CoMP environments. Since different tasks may involve varying numbers of BSs and UEs, the corresponding state and action spaces have different dimensionalities. To enable joint multi-task training, all state and action representations are zero-padded to the maximum dimensions observed across the training tasks. With a moderate number of UEs and BSs, the state space dimension can exceed 100, which makes it more challenging. Invalid action dimensions introduced by padding are masked during both training and inference, ensuring they do not affect action prediction. Following the standard offline RL setting, the collected trajectories include both successful and suboptimal interactions combined to provide diverse behavioral patterns for learning. For each one of the 36 tasks detailed in Table \ref{tab:task_config}, 2000 episodes were collected. It is also important to note that, although DTs typically require larger offline datasets than value-based offline RL methods, they are robust to low-quality data and perform well in sparse-reward environments~\cite{bhargava2024when, 10827032}.
For practical network deployments, network digital twins can facilitate offline trajectory generation under diverse operating conditions, enabling DT-based policy learning without requiring direct exploration in the operational network~\cite{11535007}.

\subsubsection{Backbone and Expert Training}
Following Stage 1 (trajectory collection), Alg.~\ref{alg:wisdom-stage12} details the first two training stages. In Stage~2 (lines 2-14), the shared backbone $f_\theta$ is trained jointly across all training tasks $\mathcal{T}^{\mathrm{train}}$: for each task, trajectories are sampled from the offline dataset $\mathcal{D}_i$ and paired with a task-identifying prompt (line~5) to form per-task minibatches $\mathcal{B}_i^M$ (lines 4-8), which are aggregated into a multi-task batch $\mathcal{B}$ (line~10). The backbone predicts actions autoregressively and is updated via masked cross-entropy loss $\mathcal{L}_{\mathrm{CE}}$ (lines 11-13), learning shared representations across all tasks. In Stage~3 (lines 15-29), the backbone is frozen and one expert $e_{\phi_c}^{(\ell)}$ is instantiated per transformer block for each task group $\mathcal{G}_c$ (line~18). Each group's expert parameters $\phi_c = \{\phi_c^{(1)},\ldots,\phi_c^{(L)}\}$ therefore consist of $L$ block-specific modules that are trained jointly, yielding $C\times L$ expert modules in total across all groups. Each expert is trained independently on its own group's batch $\mathcal{B}_c$, combined with the frozen FFN output in parallel (line~22), and only the expert parameters $\phi_c$ are updated (lines 24--26), allowing every expert to specialize on its task subset without interference from other groups. 

\subsubsection{Task Grouping}
Considered CoMP tasks vary in the number of BSs, UEs, UE velocity, and scheduler policy (ReF/PF), e.g., 5BS\_UE8\_V3\_ReF, spanning a heterogeneous range of network scales and mobility conditions. Training a single shared backbone jointly across all such tasks induces gradient conflicts, as updates favorable to one configuration can be detrimental to others.

To mitigate this, tasks are grouped by optimization behavior rather than raw configuration labels following the approach in \cite{harmoDT}. For each task $\mathcal{T}_i$, the backbone gradient $g_i = \frac{1}{B}\sum_{b=1}^{B}\nabla_{\theta}\mathcal{L}_i^{(b)}$ is computed and compared to the mean gradient $\bar{g} = \frac{1}{N}\sum_{i=1}^{N} g_i$ via the agreement vector $a_i = g_i \odot \bar{g}$. Pairwise task affinity $S_{ij} = \frac{a_i^{\top}a_j}{\|a_i\|\|a_j\|}$ is then clustered with $K$-means to partition $\{\mathcal{T}_1,\ldots,\mathcal{T}_{36}\}$ into six groups $\{\mathcal{G}_1,\ldots,\mathcal{G}_6\}$ as shown in Table \ref{tab:task_groups}, so each expert in Stage~3 specializes on tasks with aligned gradients rather than just similar configurations. 

Inspired by 
\cite{kong2025mastering}, which shows that increasing the number of experts can improve task scalability by reducing the learning burden on each parameter subset, we employ six experts in the proposed WiSDoM framework. Although larger expert counts may further improve specialization, they also increase routing complexity and training cost. This choice provides a practical trade-off between expert specialization and computational overhead while maintaining sufficient training data within each task group.

\subsubsection{Router Training and Inference} 
Alg. \ref{alg:wisdom-stage3} presents router training and evaluation. With the backbone $\theta$ and all experts $\Phi$ frozen, a router $r_{\psi}^{(\ell)}$ is introduced at each block and trained on multi-task batches. For each token, the router computes expert probabilities, selects the top-$K_{\mathrm r}$ experts, and combines their weighted outputs with the frozen FFN path (lines 6--10). As Top-$K$ selection is non-differentiable, a straight-through estimator (STE) is used (line~8): the forward pass uses sparse Top-$K_{\mathrm r}$ weights, while gradients are propagated through the dense softmax probabilities \cite{bengio2013estimating}. To mitigate expert collapse, router training also employs the auxiliary load-balancing loss used in \cite{fedus} with masked cross-entropy loss (line~13). This loss encourages a more balanced distribution of tokens across the available experts. Consequently, it prevents the router from consistently favoring a small subset of experts. For block $\ell$, it is defined as:
\begin{equation}
\mathcal{L}_{\mathrm{aux}}^{(\ell)}=
C \sum_{e=1}^{C}
f_e^{(\ell)} p_e^{(\ell)},
\end{equation}
where $f_e^{(\ell)}$ is the fraction of tokens assigned to expert $e$ and $p_e^{(\ell)}$ is its average routing probability. The auxiliary losses are averaged across all transformer blocks to obtain $\mathcal{L}_{\mathrm{aux}}$. The router is therefore optimized using:
\begin{equation}
\mathcal{L}_{\mathrm{router}}=
\mathcal{L}_{\mathrm{CE}}^{\mathrm{router}}+
\lambda_{\mathrm{aux}}\mathcal{L}_{\mathrm{aux}},
\end{equation}
while only $\psi$ is updated.
This allows the router to learn dynamic token-level expert selection without task identifiers. During evaluation (lines 16--24), sparse Top-$K_{\mathrm r}$ routing is retained. For each task $\mathcal{T}_j\in\mathcal{T}^{\mathrm{eval}}$, the model is conditioned on its best-performing prompt and a target return. Actions are then generated in closed loop from the prompt and interaction history, while the RTG and trajectory are updated after each environment interaction. An offline dataset $\mathcal{D}_i$ is assumed available for each training task, and demonstration trajectories $\mathcal{P}_i$ are available during both training and evaluation.

\begin{table}[htbp]
\centering
\caption{36 Multi-Task Configurations}
\label{tab:task_config}
\setlength{\tabcolsep}{3.5pt}
\footnotesize
\begin{tabular}{cccccc}
\toprule
\textbf{Layout} & \textbf{\#UE} & \textbf{State Dim.} & \textbf{Map Size (m)} & \textbf{Vel. (m/s)} & \textbf{Sched.} \\
\midrule
\multirow{2}{*}{3BS} & 3 & 21 & \multirow{2}{*}{$120\times107$} & \multirow{2}{*}{0.5, 1.5, 3.0} & \multirow{2}{*}{PF, ReF} \\
                     & 5 & 35 &                                  &                                 &                          \\
\midrule
\multirow{2}{*}{4BS} & 4 & 36 & \multirow{2}{*}{$120\times120$} & \multirow{2}{*}{0.5, 1.5, 3.0} & \multirow{2}{*}{PF, ReF} \\
                     & 5 & 45 &                                  &                                 &                          \\
\midrule
\multirow{2}{*}{5BS} & 6 & 66 & \multirow{2}{*}{$210\times190$} & \multirow{2}{*}{0.5, 1.5, 3.0} & \multirow{2}{*}{PF, ReF} \\
                     & 8 & 88 &                                  &                                 &                          \\
\bottomrule
\end{tabular}
\end{table}
\begin{table}[h!]
\centering
\caption{Simulation Parameters and Hyperparameters}
\label{tab:simulation_parameters}
\setlength{\tabcolsep}{5pt}
\footnotesize
\renewcommand{\arraystretch}{1.08}
\begin{tabular}{>{\raggedright\arraybackslash}p{0.43\columnwidth}
                >{\raggedright\arraybackslash}p{0.43\columnwidth}}
\toprule
\textbf{Parameter} & \textbf{Value} \\
\midrule

\multicolumn{2}{c}{\textbf{CoMP Environment Parameters}} \\
\midrule
$f_c$ (GHz), BW (MHz)                 & 2.5, 9 \\
UE Velocity                    & [0.5,1.5,3] m/s \\
Scheduler (s)                      & PF, ReF \\
$P_{Tx}$ (dBm), $N_0$ (dBm/Hz)                & 30, -90 \\
$\rho_{min}$ (dB)               & -77\\
Cell tower and UE height (m)               & 50 \& 1.5\\
Cell-to-cell distance (m) & 100 \\
$(U_{\min}, U_{\max})$        & $(-10, 10)$ \\
$(w_1,w_2,w_3)$          & $(10,0,10)$ \\
Episode length & 100\\
\midrule
\multicolumn{2}{c}{\textbf{Backbone, Expert and Router Parameters}} \\
\midrule
n\_heads, n\_blocks         & 4, 6\\
$K$, $K^\star$  & 20, 5 \\
per-task batch\_size     & 8  \\
reward\_scale & 50\\
optimizer       & AdamW  \\
learning\_rate,weight\_decay          & $10^{-4}$ ,$10^{-4}$ \\
activation function            & ReLU \& GELU \\
Number of Experts & 6 \\
N1, N2, N3 & 400k, 400k, 150k \\
$\lambda_{\mathrm{aux}}$ & 0.01\\
hidden\_dim(s)  & 256,448,1024 \\
MoE Type & Token-level Top-K (1-2)\\
\bottomrule
\end{tabular}
\end{table}
\vspace{-10pt}
\section{Simulation Settings and Numerical Results}
\label{sec:sim_settings_and_results}
In this section, we first summarize multi-task scenarios, simulation settings, and model hyperparameters and then present the comprehensive numerical results. 
In our study, we construct a set of multi-task learning scenarios based on the open-source system-level lightweight simulation platform, namely mobile-env \cite{9789886}. We consider a total of 36 tasks, as shown in Table \ref{tab:task_config}, varying the number of BSs, UEs, UE velocities, and scheduler policies. Also, CoMP environment simulation parameters and model hyperparameters are detailed in Table \ref{tab:simulation_parameters}. All models were trained on a single NVIDIA RTX 5090 GPU (32 GB VRAM, ~105 TFLOPS FP32). 

\vspace{-10pt}
\subsection{Baselines}
\label{sec:baselines}
We compare WiSDoM against four baseline categories, all evaluated on the identical 36-task suite and held-out unseen tasks (e.g., an extra 8 tasks) using the same closed-loop rollout (50 episodes/task).

\begin{itemize}
    \item \textbf{Heuristic:} Following \cite{schneider2023deepcomp,article_heuristic}, a UE $u_j$ associates with every BS $b_i$ whose SINR falls within a threshold fraction $\epsilon$ of the best available SINR at that UE, i.e.\ $\text{SINR}_{i,j}(t) \geq \epsilon \cdot \text{SINR}_j^{\max}(t)$. We sweep $\epsilon \in \{0, 0.5, 1\}$: at $\epsilon = 1$ only the strongest cell is retained, recovering the single-cell association, while $\epsilon = 0$ connects to all cells if thresholds are satisfied. The heuristic has complexity $\mathcal{O}(UB\log B)$ due to per-UE BS sorting. While computationally efficient, its performance is sensitive to $\epsilon$, whose optimal value varies across network configurations and requires manual tuning.
    \item \textbf{PPO (single-task DRL):} PPO models \cite{schulman2017ppo} are trained independently per task (36 separate policies); upper bound on task-specific specialization, no cross-task generalization. Default PPO hyperparameters in \cite{schneider2023deepcomp} are used.  
    \item \textbf{PromptDT (multi-task DT):} PromptDT trained jointly across all 36 tasks, evaluated at 5M/15M/75M parameter scales to isolate the effect of backbone capacity alone \cite{Temiz2026Generalizable}. Larger models were obtained by increasing only the transformer hidden dimension while keeping the architecture (e.g., number of blocks and attention heads) unchanged, as outlined in Table \ref{tab:simulation_parameters}. Training logic follows backbone training in Alg. \ref{alg:wisdom-stage12}.
    \item \textbf{WiSDoM (proposed):} MoE-augmented DT with Top-1/Top-2 token-level routing over per task-group experts. In the remainder, Top-1 router based architecture called as WiSDoM-75M-A27M and Top-2 router based architecture called as WiSDoM-75M-A37M, showing a total of 75M parameters and active numbers of parameters 27M and 37M, respectively. PromptDT-15M is used as the backbone of the architecture. 
\end{itemize}

\begin{figure}[h!]
    \centering
    \includegraphics[width=1\columnwidth
    ]{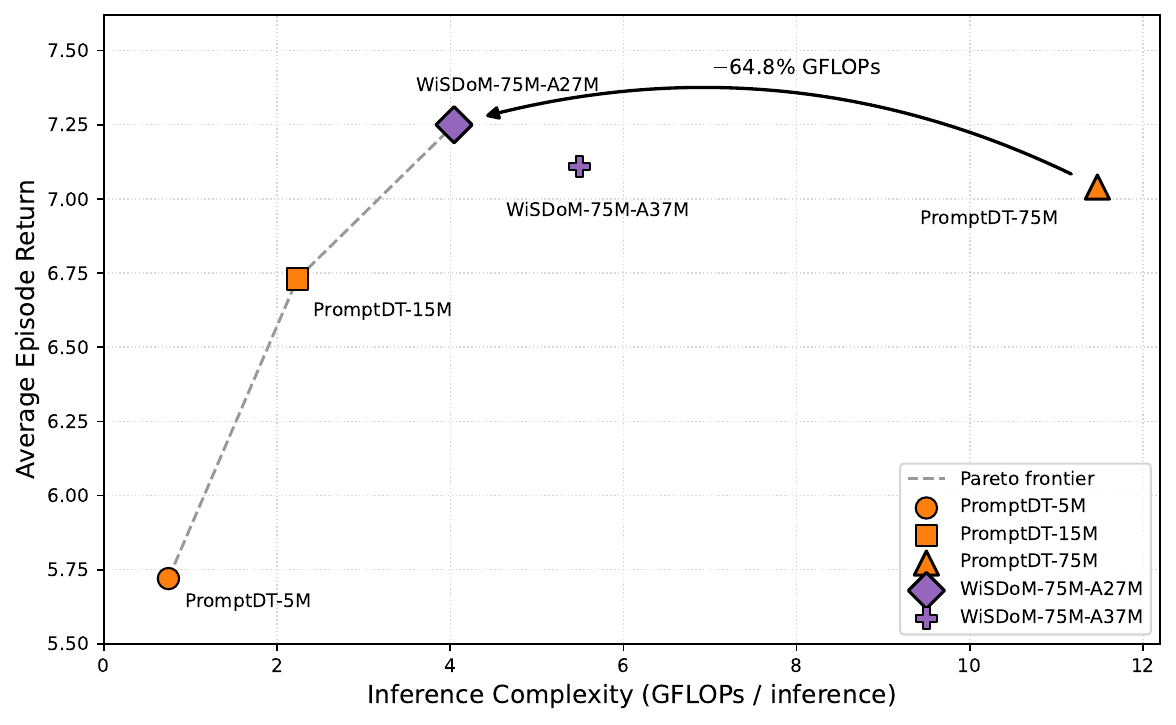}
    \caption{Average Episode Return vs. Inference Complexity}
    \label{fig:pareto}
    \vspace{-10pt}
\end{figure}

\subsection{Results:}
We compare the methods along five axes: (i) RL performance, (ii) network performance (QoE, rate outage probability, and active UEs), (iii) task-wise comparison (consistency across 36 tasks under a shared model), (iv) unseen-task performance (generalization to held-out configurations), and (v) complexity (parameter count, active parameters per forward pass, and inference latency).

\subsubsection{RL Performance}
We first compare the proposed WiSDoM models against PromptDT multi-tasking baselines on RL performance versus inference complexity across all evaluation tasks as shown in Fig. \ref{fig:pareto}. Increasing the PromptDT model size from 5M to 75M parameters consistently improves the average episode return from 5.72 to 7.04, indicating improved multi-task learning capacity with model scaling. The proposed sparse MoE architecture further improves performance while activating only a subset of experts during inference. In particular, WiSDoM-75M-A27M with Top-1 routing achieves the highest average episode return of 7.25, outperforming the dense PromptDT-75M (7.04), while WiSDoM-75M-A37M with Top-2 routing achieves 7.11 despite both using significantly fewer active parameters (e.g., up to 3x) during inference.

\begin{figure}[h!]
    \centering
    \includegraphics[width=1\columnwidth, height=4cm]{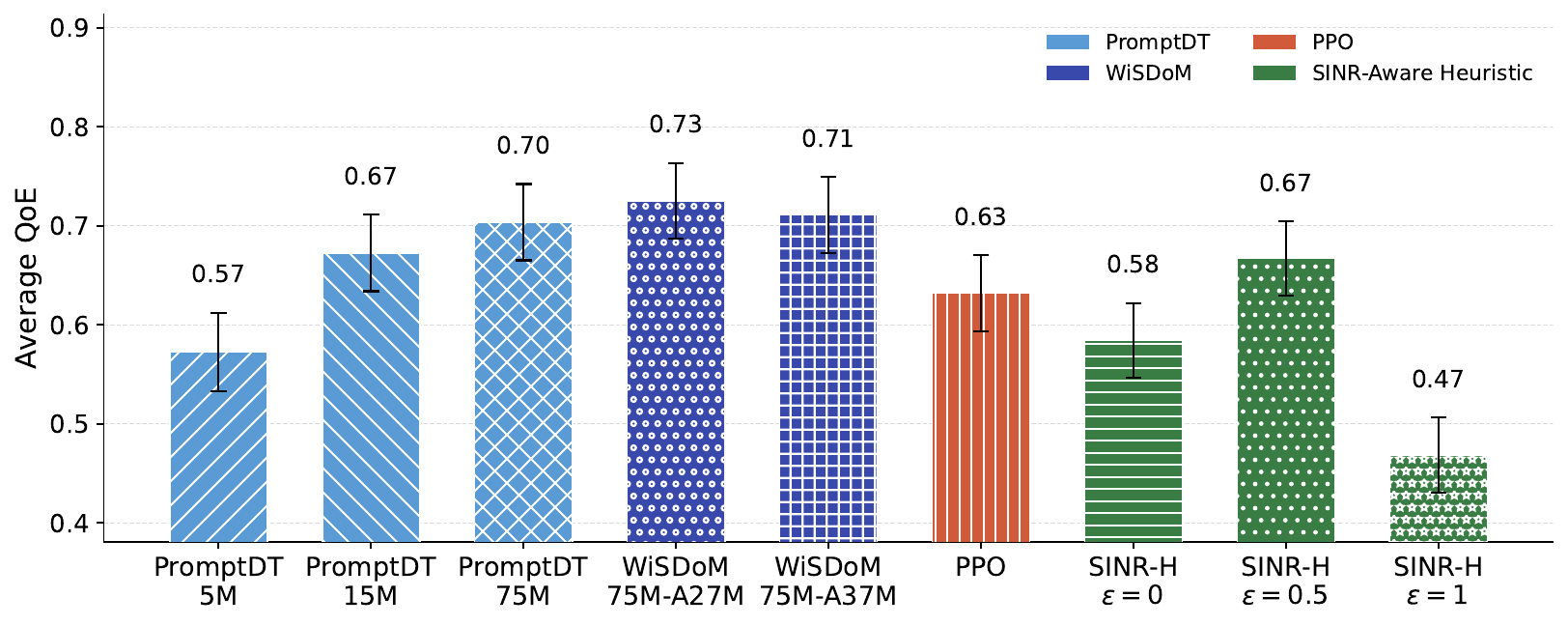}
    \caption{Average QoE Across 36 Tasks}
    \label{fig:reward} 
\end{figure} 

\begin{figure*}[h]
    \centering

    \begin{subfigure}[t]{0.32\textwidth}
        \centering
        \includegraphics[width=\linewidth]{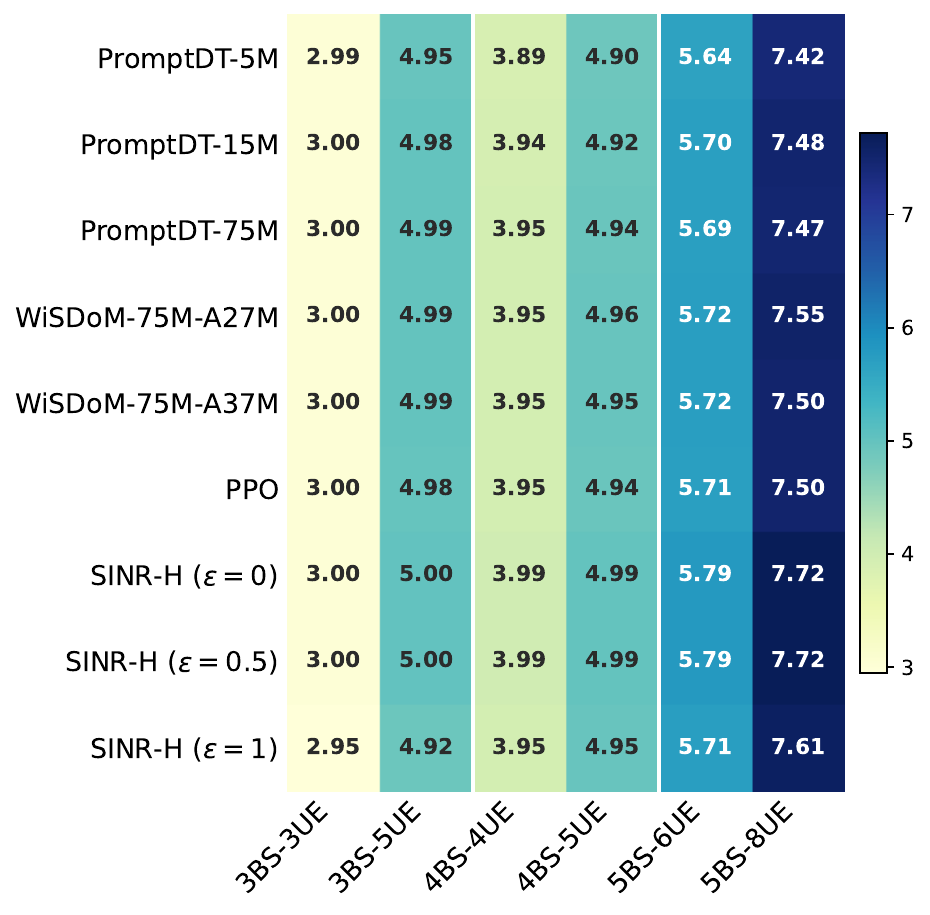}
        \caption{Average number of connected UEs.}
        \label{fig:heatmap_connected_ues}
    \end{subfigure}
    \hfill
    \begin{subfigure}[t]{0.32\textwidth}
        \centering
        \includegraphics[width=\linewidth]{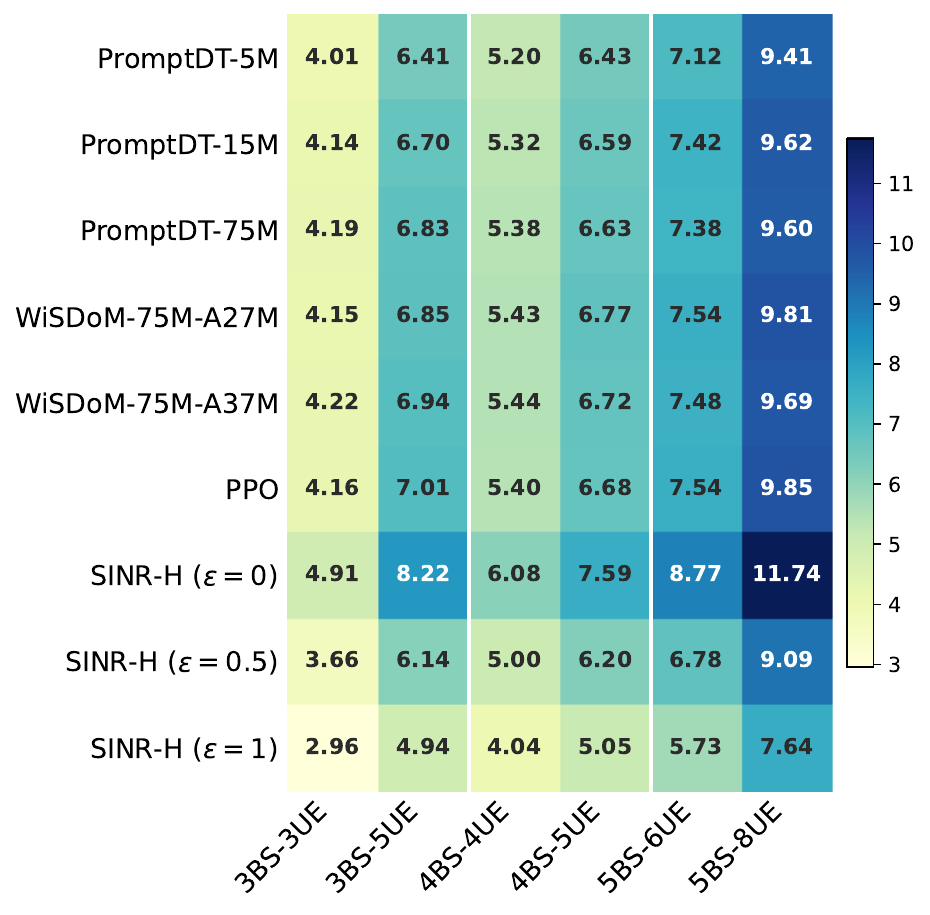}
        \caption{Average number of established links.}
        \label{fig:heatmap_established_links}
    \end{subfigure}
    \hfill
    \begin{subfigure}[t]{0.32\textwidth}
        \centering
        \includegraphics[width=\linewidth]{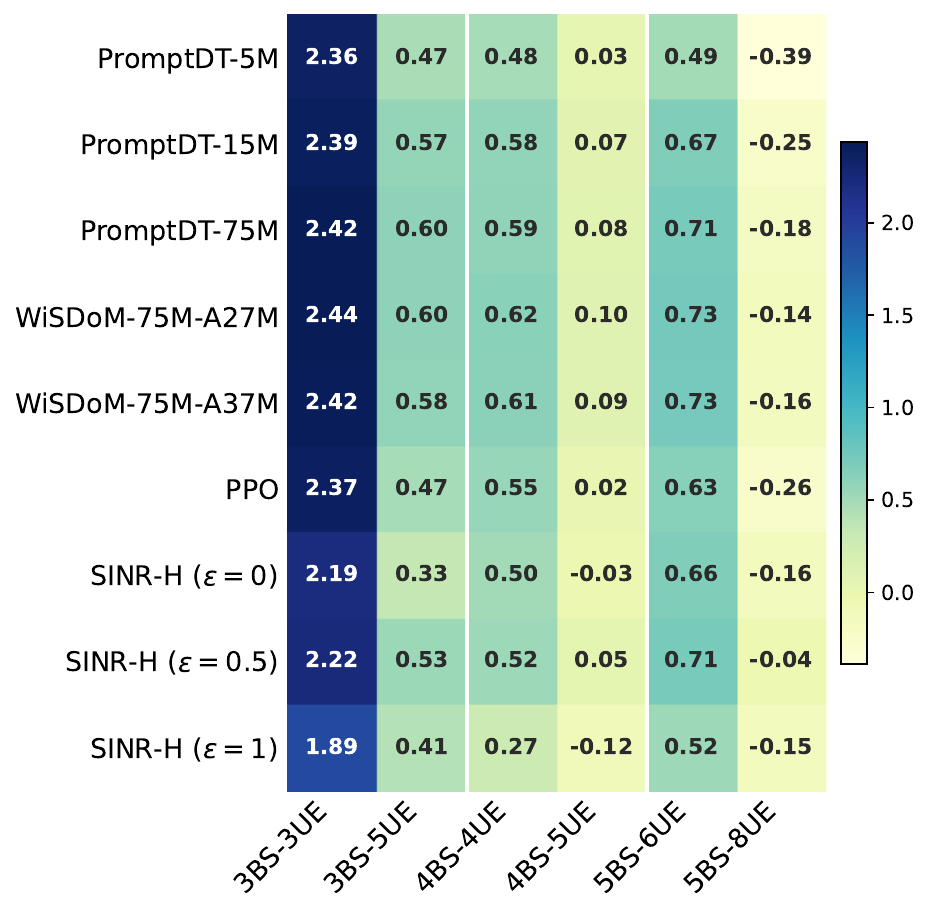}
        \caption{Average QoE.}
        \label{fig:heatmap_qoe}
    \end{subfigure}

    \caption{Performance comparison across network configurations in terms of the average number of connected UEs, established links, and QoE. }
    \label{fig:network_configuration_heatmaps}
\end{figure*}
\begin{figure}[h!]
    \centering
    \includegraphics[width=0.85\columnwidth]{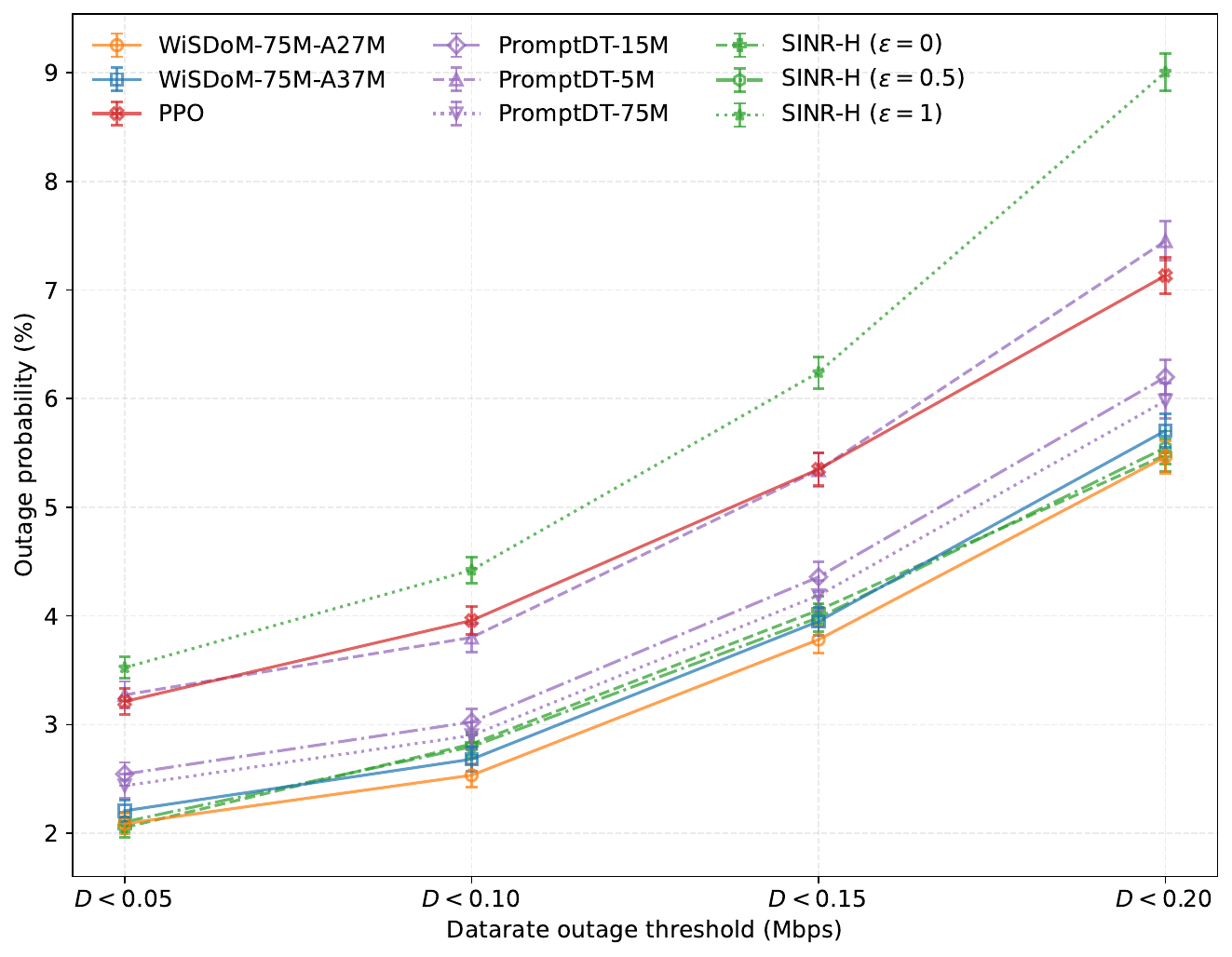}
    \caption{Rate Outage Probability (\%)}
    \label{fig:rateoutageall} 
    \vspace{-15pt}
\end{figure} 
Compared with PPO, WiSDoM improves users' QoE from 0.63 to 0.73 (15.87\%) and outperforms all heuristic schedulers up to 55\%. The SINR-H with $\epsilon=0.5$ achieves better QoE (0.67) compared to others, and the remaining configurations perform considerably worse. These results demonstrate that sparse expert routing preserves the representational capacity of a dense transformer while substantially improving inference efficiency. 
Furthermore, WiSDoM learns a single unified policy that operates across all 36 network configurations, whereas PPO requires 36 independently trained policies. This significantly simplifies AI deployment in next-generation RANs, especially when hierarchical RICs operate under diverse network conditions \cite{10766614}. Instead of maintaining, selecting, and updating dozens of task-specific models, operators can deploy and manage a single policy, reducing model lifecycle management, software updates, and operational complexity while facilitating continual adaptation to evolving network scenarios.

\begin{figure*}[t!]
    \centering
    \includegraphics[width=2\columnwidth]{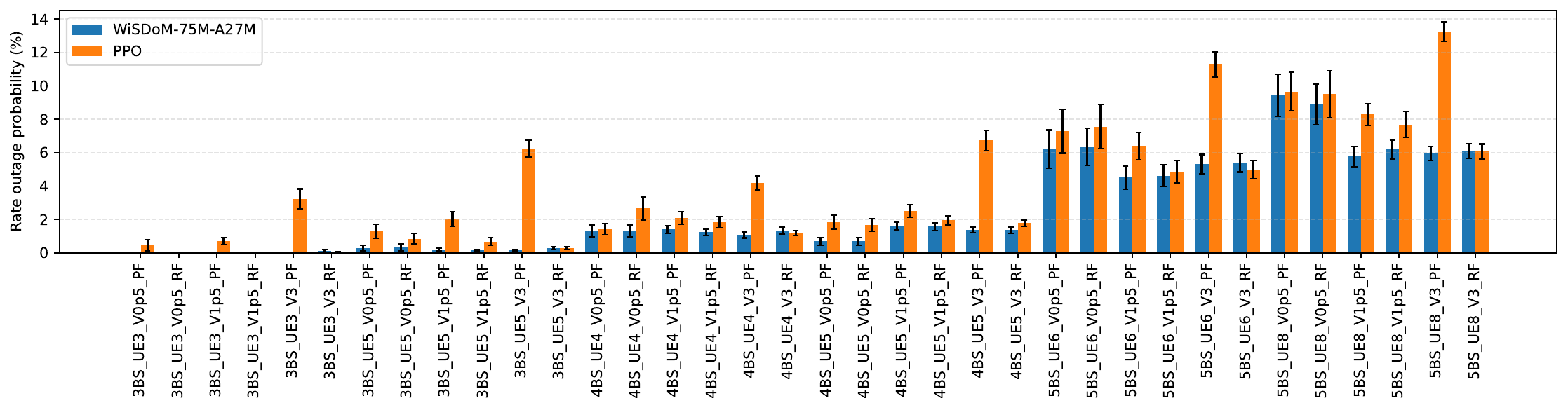}
    \caption{Task-wise rate outage probability of WiSDoM and PPO across 36 tasks.}
    \label{fig:taskwise} 
    \vspace{-10pt}
\end{figure*} 

\subsubsection{Network Performance}
To better understand the QoE performance of the evaluated policies, Figs. \ref{fig:heatmap_connected_ues}, \ref{fig:heatmap_established_links}, and \ref{fig:heatmap_qoe} provide complementary insights into their network-level behavior across six configurations, ranging from 3BS-3UE to 5BS-8UE. Each configuration reports the average over six tasks corresponding to two scheduling policies and three UE mobility settings. As shown in Fig. \ref{fig:heatmap_connected_ues}, the average number of connected UEs is nearly identical across the learning-based approaches. Similarly, Fig. \ref{fig:heatmap_established_links} shows relatively small differences in the number of established links among PromptDT, WiSDoM, and PPO. These observations are important for interpreting the QoE results: the performance differences are not simply explained by serving more UEs or establishing substantially more CoMP links. Instead, the results suggest that how the available BS associations and radio resources are utilized plays a more important role in the resulting QoE. This is particularly evident when comparing the learning-based approaches with the SINR-aware heuristics. The SINR-H policy with $\epsilon=0$ generally establishes considerably more links than the learned policies since it associates each UE with all BSs whose SINR exceeds the threshold $\rho_{ij}$ without considering BS load balancing or long-term resource allocation, leading to excessive CoMP links that do not translate into improved QoE. For example, in the 3BS-5UE configurations, it establishes 8.22 links on average, compared with 6.85 for WiSDoM-75M-A27M, while their corresponding QoE values are $0.33$ and $0.60$, respectively. In contrast, SINR-H with $\epsilon=1$ generally yields the lowest QoE, as it associates each UE only with its strongest BS and therefore does not exploit the potential gains of CoMP coordination. It demonstrates the limitations of purely threshold-based association. Fig. \ref{fig:heatmap_qoe} further shows that WiSDoM-75M-A27M achieves the highest or joint-highest average QoE in five of the six network configurations. DT-based models show overall better performance compared to single-task RL and heuristic baselines. For instance, in the 3BS-5UE configuration, WiSDoM achieves a QoE of 0.60, corresponding to a 27.7\% improvement over PPO (0.47). As the number of BSs and UEs increases, the dimensionality of the joint state and action spaces also grows substantially, making policy learning increasingly challenging for task-specific PPO agents. These observations indicate that while conventional DRL methods perform competitively in relatively simple environments, their performance deteriorates in larger environments, and this can be explained as the state space becomes increasingly complex (e.g., 88 in the largest case as shown in Table \ref{tab:task_config}).

Beyond cumulative reward and average QoE, we evaluate service reliability using the rate outage probability, defined as $P_{\mathrm{out}}=\Pr(D_j < \tau)$, where $D_j$ denotes the aggregated UE DL data rate and $\tau \in {0.05,0.10,0.15,0.20}$ Mbps. As shown in Fig.~\ref{fig:rateoutageall}, outage probability increases with $\tau$ for all methods, while WiSDoM consistently remains among the best-performing approaches. Across the four thresholds, WiSDoM-75M-A27M reduces outage probability by $8.6$--$14.3\%$ relative to PromptDT-75M and by $23.4$--$36.0\%$ relative to PPO. WiSDoM also remains competitive with the SINR-H baselines. Although SINR-H with $\epsilon=0$ achieves a marginally lower outage at $\tau=0.05$ Mbps ($2.05\%$ versus $2.09\%$), WiSDoM achieves lower outage at $\tau=0.10$ and $\tau=0.15$ Mbps and nearly identical performance at $\tau=0.20$ Mbps. Also, it is important to note that this reliability performance comes with the cost of overloaded BSs and degraded QoE for SINR-H approaches, as illustrated in Figs. \ref{fig:heatmap_established_links} and \ref{fig:heatmap_qoe}.  

\begin{figure*}[h!]
    \centering
    \includegraphics[width=2\columnwidth,]{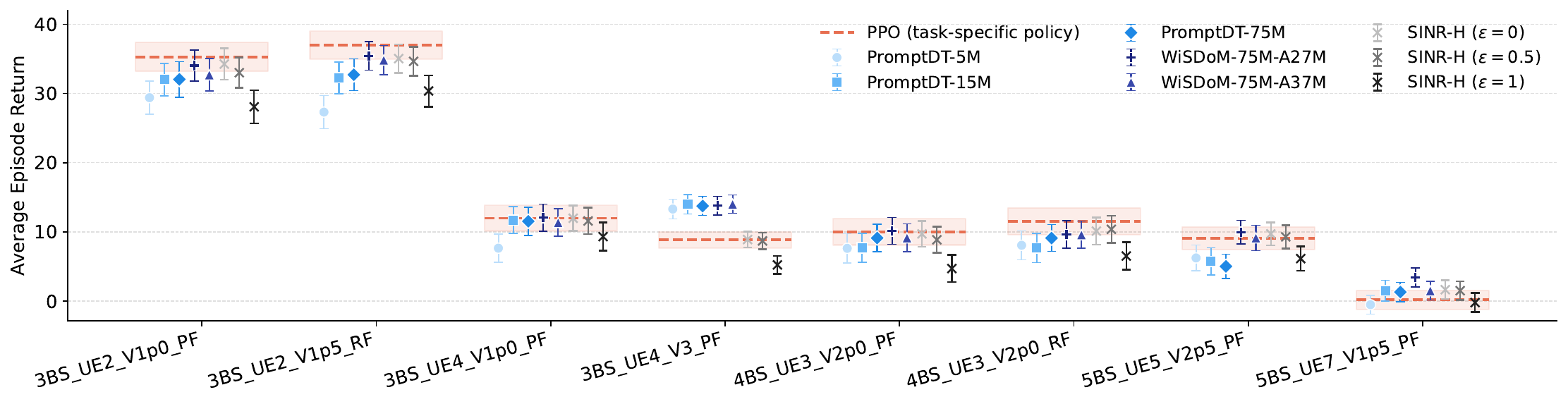}
    \caption{Unseen Task Performance Results}
    \label{fig:unseen}
    \vspace{-10pt}
\end{figure*} 
\subsubsection{Task-Wise Performance}
To further examine the reliability gains at a finer granularity, Fig.~\ref{fig:taskwise} compares the task-wise rate outage probabilities of WiSDoM-75M-A27M and PPO across all 36 evaluation tasks. WiSDoM achieves a lower outage probability than PPO in 32 out of
36 tasks (88.89\%), including all 18 tasks with PF scheduling.
The advantage is particularly pronounced under high mobility with PF
scheduling. At $3$~m/s, WiSDoM reduces the outage probability from
6.22\% to 0.16\% in the 3BS-5UE configuration. Similarly, for the larger 5BS-6UE and 5BS-8UE
configurations, the outage probability decreases from 11.29\% to 5.31\%
and from 13.23\% to 5.96\%, corresponding to reductions of approximately
52.9\% and 55.0\%, respectively. In contrast, the performance gap under ReF scheduling is generally smaller than under PF scheduling, although WiSDoM still provides clear improvements in several tasks. These results indicate that
WiSDoM provides more reliable data-rate performance, particularly under dynamic and increasingly complex network conditions.

\subsubsection{Unseen Task Performance}

Sequence-based decision modelling enables adaptation to unseen tasks through in-context conditioning. We evaluate the DT-based models on eight unseen network configurations against task-specific PPO policies and SINR-aware heuristics, as shown in Fig.~\ref{fig:unseen}. The DT-based models use demonstration trajectories of length $K^{\star}$ as prompts without target-task parameter updates or fine-tuning, whereas a separate PPO policy is trained for each unseen task.
WiSDoM demonstrates strong transfer using a single shared pretrained policy, outperforming task-specific PPO on five of the eight unseen tasks. For example, on the 3BS-4UE configuration with $v=3.0$~m/s and PF scheduling, WiSDoM-75M-A27M achieves an average return of 13.8 compared with 8.8 for PPO. On the challenging 5BS-7UE configuration, WiSDoM achieves the highest return of 3.4, compared with 0.2 for PPO and 1.3 for PromptDT-75M. Even on configurations where PPO performs best, WiSDoM remains competitive.
These results demonstrate that WiSDoM can reuse knowledge acquired across previously observed environments to adapt to new network configurations through few-shot conditioning, reducing the need to train and maintain a dedicated policy for each task.

\begin{figure}[h!]
    \centering
    \includegraphics[width=0.95\columnwidth]{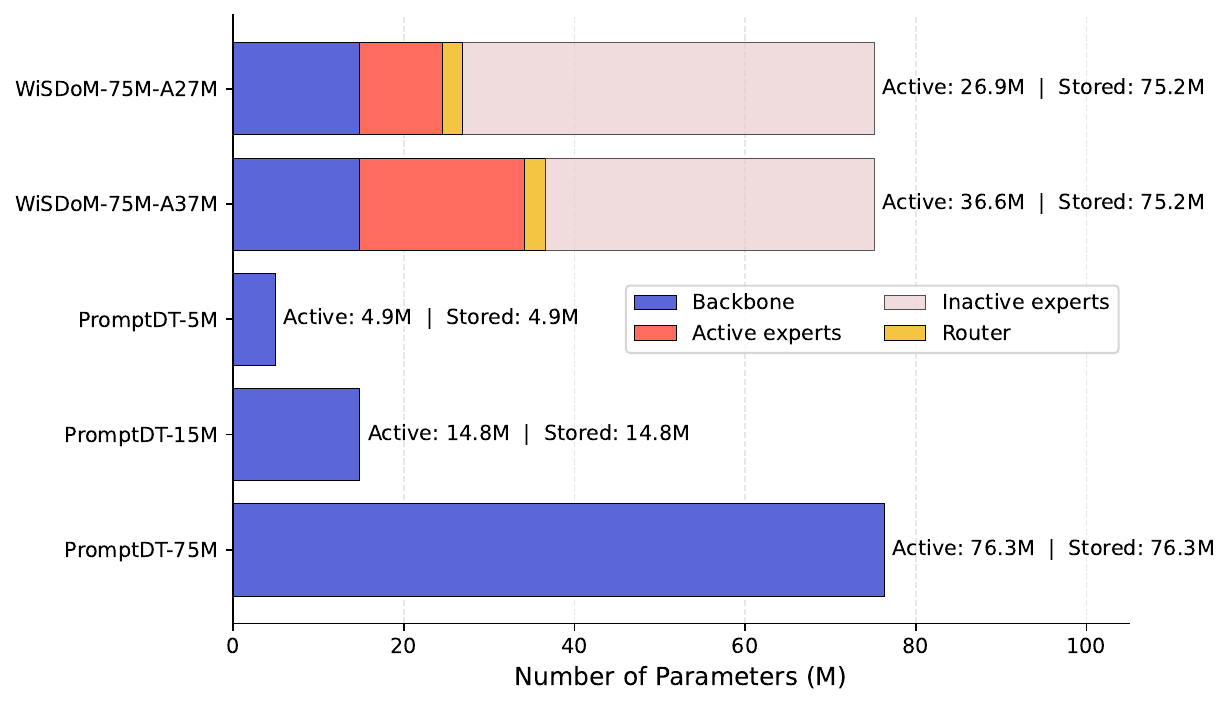}
    \caption{Active and Stored Model Parameters}
    \label{fig:actparam} 
    \vspace{-10pt}
\end{figure} 
\vspace{-10pt}
\subsection{Complexity Analysis and Ablation Study}

\subsubsection{Complexity Analysis} 
While WiSDoM stores the same total number of parameters as the dense PromptDT-75M, only the backbone, router, and selected experts (e.g., Top-1 or Top-2) are activated during inference. Hence, WiSDoM-75M-A27M activates only 26.9M parameters, and WiSDoM-75M-A37M activates 36.6M compared with 76.3 for the dense PromptDT-75M.

This reduction directly translates into lower computational complexity. As summarized in Table~\ref{tab:compute_memory}, PromptDT-75M requires 11.47 GFLOPs per inference step as measured empirically (i.e., profiling). In contrast, WiSDoM-75M-A27M reduces the computational cost to 4.044 GFLOPs (35\% of the dense model), and WiSDoM-75M-A37M requires 5.490 GFLOPs (48\% of the dense model). Thus, WiSDoM substantially reduces the arithmetic computation required for inference while maintaining and, in many cases, improving task performance, demonstrating an attractive performance-efficiency tradeoff for practical deployments such as resource-constrained near real-time RIC. As demonstrated, sparse activation can further improve the energy efficiency of larger models by reducing the amount of computation performed during inference, thereby enhancing test-time scalability and lowering resource consumption \cite{xu2026moesurvey}. Furthermore, despite their large parameter counts, all DT-based models achieve per-decision inference latencies of 8-20~ms on a consumer-grade GPU with 16~GB of VRAM. Further optimizations can lower this inference time on various hardware variants in real networks \cite{yang2026freetokenefficientedgenativemoe, 11527015}.

\begin{table}[h!]
\centering
\caption{Computational complexity comparison}
\label{tab:compute_memory}
\setlength{\tabcolsep}{3.5pt}
\footnotesize
\begin{tabular}{lccc}
\toprule
\textbf{Model} &
\shortstack{\textbf{FP16}-\textbf{Memory}} &
\shortstack{\textbf{GFLOPs}\textbf{/Inf.}} &
\shortstack{\textbf{GFLOPs}\textbf{/Token}} \\
\midrule
WiSDoM-75M-A27M & 150.3 MB  & 4.044  & 0.054 \\
WiSDoM-75M-A37M & 150.3 MB   & 5.490  & 0.073 \\
PromptDT-5M      & 9.8 MB   & 0.745  & 0.010 \\
PromptDT-15M     & 29.6 MB  & 2.233  & 0.030 \\
PromptDT-75M     & 152.6 MB & 11.474 & 0.153 \\
\bottomrule
\end{tabular}
\end{table}
\vspace{-10pt}
\subsection{Ablation Study on Expert Specialization and Routing}
\begin{figure}[h!]
    \centering
    \includegraphics[width=0.8\columnwidth]{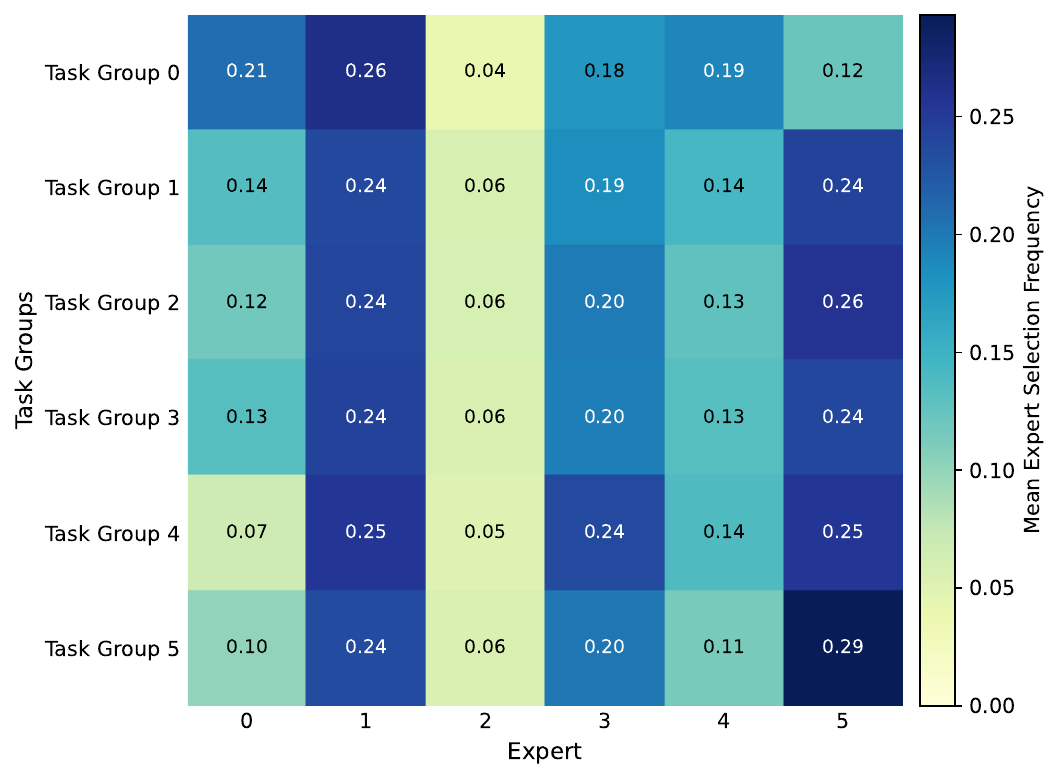}
    \caption{Mean expert selection frequency for each task group}
    \label{fig:expert_util}
    \vspace{-10pt}
\end{figure}

Similar to the expert specialization and router strategy analysis in \cite{jiang2024mixtral}, we investigate whether experts specialize in particular task groups or learn shared knowledge across tasks, and interestingly, we observe both behaviors. Fig.~\ref{fig:expert_util} shows the mean expert selection distribution for each task group defined in Table~\ref{tab:task_groups}, computed across all six transformer blocks and each row sums up to 1. Experts 0 and 5 exhibit clear task-dependent specialization, with higher utilization for the task groups on which they were trained, i.e., Groups 0 and 5, respectively. Furthermore, Expert 0 is primarily associated with smaller network configurations, whereas Expert 5 is more frequently selected for challenging high-mobility and larger-network tasks. In contrast, Experts 1, 3, and 4 behave as generalists, maintaining relatively consistent utilization across task groups. Such behavior is expected and further supports the coexistence of shared and specialized knowledge, consistent with recent frontier MoE architectures that employ shared experts alongside dynamically routed experts \cite{deepseekai2025deepseekv3technicalreport}.
In addition, although Expert 2 is less frequently selected, no expert collapse is observed, owing to the auxiliary load-balancing loss. We further observe distinct expert-selection patterns across transformer blocks (i.e., layers). 

\begin{table}[t]
\centering
\caption{Task groups}
\label{tab:task_groups}
\footnotesize
\resizebox{\columnwidth}{!}{
\begin{tabular}{c|l}
\hline
\textbf{Group} & \textbf{Tasks} \\ \hline
0 &
\begin{tabular}[c]{@{}l@{}}
4BS-UE4-V0.5-ReF, 4BS-UE4-V0.5-PF, 3BS-UE3-V0.5-PF, 3BS-UE5-V0.5-PF,\\
4BS-UE5-V0.5-ReF, 3BS-UE3-V0.5-ReF, 4BS-UE5-V0.5-PF, 3BS-UE5-V0.5-ReF,\\
3BS-UE3-V1.5-PF, 4BS-UE4-V1.5-ReF
\end{tabular} \\ \hline
1 &
\begin{tabular}[c]{@{}l@{}}
4BS-UE4-V3-ReF, 4BS-UE5-V3-ReF, 3BS-UE5-V3-ReF,\\
3BS-UE3-V3-ReF, 5BS-UE6-V3-ReF
\end{tabular} \\ \hline
2 &
\begin{tabular}[c]{@{}l@{}}
3BS-UE5-V1.5-PF, 4BS-UE5-V1.5-PF, 5BS-UE6-V1.5-PF,
4BS-UE4-V1.5-PF
\end{tabular} \\ \hline
3 &
\begin{tabular}[c]{@{}l@{}}
3BS-UE5-V1.5-ReF, 3BS-UE3-V1.5-ReF, 4BS-UE5-V1.5-ReF,\\
5BS-UE6-V1.5-ReF, 5BS-UE8-V1.5-PF
\end{tabular} \\ \hline
4 &
\begin{tabular}[c]{@{}l@{}}
5BS-UE6-V0.5-ReF, 5BS-UE6-V0.5-PF,\\
5BS-UE8-V0.5-ReF, 5BS-UE8-V0.5-PF
\end{tabular} \\ \hline
5 &
\begin{tabular}[c]{@{}l@{}}
3BS-UE5-V3-PF, 4BS-UE5-V3-PF, 3BS-UE3-V3-PF, 4BS-UE4-V3-PF,\\
5BS-UE6-V3-PF, 5BS-UE8-V3-PF, 5BS-UE8-V3-ReF, 5BS-UE8-V1.5-ReF
\end{tabular} \\ \hline
\end{tabular}}
\end{table}

\section{CONCLUSION}
\label{sec:conclusion}

This paper proposed WiSDoM, a MoE-DT-based multi-task learning framework for multi-cell selection in CoMP that addresses key limitations of conventional DRL, including limited scalability, and the need to train separate policies for different network configurations. By integrating prompt-based task conditioning with sparse expert routing, WiSDoM enables a single model to specialize across heterogeneous wireless environments while maintaining efficient conditional computation through sparse expert activation.
Extensive experiments over 36 CoMP tasks show that WiSDoM consistently outperforms heuristic, single-task, and conventional multi-task baselines in QoE and rate outage probability, achieving up to 55\% higher QoE, with its advantage most pronounced on harder task configurations where load-blind and fixed-threshold-based heuristics degrade.
In addition, its sparse MoE architecture activates as few as one-third of the parameters of an equivalent-sized dense Transformer during inference without compromising performance. The proposed framework also exhibits strong scalability with increasing model capacity, effectively mitigates task interference through expert specialization, and generalizes well to previously unseen network configurations, enabling robust few-shot adaptation without retraining or fine-tuning.
Future work will investigate MoE architecture in more detail, including auxiliary-loss-free balancing strategies, as well as robustness beyond the trained model's dimensional bounds and applications to O-RAN architectures.
\vspace{-15pt}
\section*{ACKNOWLEDGMENT}
This work has been supported by the NSERC Canada Research Chairs Program.

\renewcommand\refname{References}
\addcontentsline{toc}{section}{References}
\bibliographystyle{IEEEtran}
\bibliography{ref}

@inproceedings{xu2022prompt,
        title={Prompting Decision Transformer for Few-Shot Policy Generalization},
        author={Xu, Mengdi and Shen, Yikang and Zhang, Shun and Lu, Yuchen 
        and Zhao, Ding and Tenenbaum, B. Joshua and Gan, Chuang},
        booktitle={ICML},
        year={2022}
    }

@article{fedus,
author = {Fedus, William and Zoph, Barret and Shazeer, Noam},
title = {Switch transformers: scaling to trillion parameter models with simple and efficient sparsity},
year = {2022},
issue_date = {January 2022},
publisher = {JMLR.org},
volume = {23},
number = {1},
issn = {1532-4435},
journal = {J. Mach. Learn. Res.},
month = jan,
articleno = {120},
numpages = {39}
}

@misc{deepseekai2025deepseekv3technicalreport,
      title={DeepSeek-V3 Technical Report}, 
      author={DeepSeek-AI and Aixin Liu and Bei Feng and others},
      year={2025},
      eprint={2412.19437},
      archivePrefix={arXiv},
      primaryClass={cs.CL},
      url={https://arxiv.org/abs/2412.19437}, 
}

@ARTICLE{10735366,
  author={Ghassemi, Mohammad and Zhang, Han and Afana, Ali and Sediq, Akram Bin and Erol-Kantarci, Melike},
  journal={IEEE Networking Letters}, 
  title={Multi-Modal Transformer and Reinforcement Learning-Based Beam Management}, 
  year={2024},
  volume={6},
  number={4},
  pages={222-226},
  doi={10.1109/LNET.2024.3486260}}

@INPROCEEDINGS{10827032,
  author={Li, Jun and Zhu, Yintao and Xia, Pengcheng and Ma, Ting and Zhang, Jie and Shi, Long and Zeng, Zhenping and Jin, Shi},
  booktitle={2024 IEEE WCSP}, 
  title={Prompt Decision Transformer Based Policy Generalization for Base Station Energy Saving}, 
  year={2024},
  volume={},
  number={},
  pages={1204-1209},
  doi={10.1109/WCSP62071.2024.10827032}}

@ARTICLE{11080254,
  author={Zhou, Tailin and others},
  journal={IEEE TWC}, 
  title={Federated Prompt-based Decision Transformer for Resource Allocation of Customized VR Streaming in Mobile Edge Computing}, 
  year={2025},
  volume={},
  number={},
  pages={1-1},
  doi={10.1109/TWC.2025.3586371}}

@inproceedings{NEURIPS2021_7f489f64,
  title     = {Decision Transformer: Reinforcement Learning via Sequence Modeling},
  author    = {Chen, Lili and others},
  booktitle = {NeurIPS},
  year      = {2021}
}

@INPROCEEDINGS{11432277,
  author={Huang, Weijun and Tham, Chen-Khong},
  booktitle={2025 IEEE GLOBECOM}, 
  title={Transformer-based Reinforcement Learning for Base Station Selection}, 
  year={2025},
  volume={},
  number={},
  pages={2342-2347},
  doi={10.1109/GLOBECOM59602.2025.11432277}}

@INPROCEEDINGS{10682015,
  author={Erak, Omar and Alhussein, Omar and Naser, Shimaa and Alabbasi, Nouf and Mi, De and Muhaidat, Sami},
  booktitle={2024 IEEE/CIC ICCC}, 
  title={Large Language Model-Driven Curriculum Design for Mobile Networks}, 
  year={2024},
  volume={},
  number={},
  pages={179-184},
  doi={10.1109/ICCC62479.2024.10682015}}

@ARTICLE{10699421,
  author={Salehi, Shavbo and Zhou, Hao and Elsayed, Medhat and Bavand, Majid and Gaigalas, Raimundas and Ozcan, Yigit and Erol-Kantarci, Melike},
  journal={IEEE TMLCN}, 
  title={Smart Jamming Attack and Mitigation on Deep Transfer Reinforcement Learning Enabled Resource Allocation for Network Slicing}, 
  year={2024},
  volume={2},
  number={},
  pages={1492-1508},
  doi={10.1109/TMLCN.2024.3470760}}

@ARTICLE{11021485,
  author={Lu, Chi and Ni, Yiyang and Wang, Zhe and Shi, Xiaoli and Li, Jun and Jin, Shi},
  journal={IEEE Wireless Communications Letters}, 
  title={Attention-Enhanced Prompt Decision Transformers for AAV-Assisted Communications With AoI}, 
  year={2025},
  volume={14},
  number={8},
  pages={2576-2580},
  doi={10.1109/LWC.2025.3575757}}

@ARTICLE{10839243,
  author={Zhang, Jie and Li, Jun and Wang, Zhe and Shi, Long and Jin, Shi and Chen, Wen and Poor, H. Vincent},
  journal={IEEE Wirel. Commun.}, 
  title={Decision Transformers For Wireless Communications: A New Paradigm Of Resource Management}, 
  year={2025},
  volume={32},
  number={2},
  pages={180-186},
  doi={10.1109/MWC.007.2400124}}

@misc{radford2019language,
  title={Language Models are Unsupervised Multitask Learners},
  author={Radford, Alec and Wu, Jeff and Child, Rewon and Luan, David and Amodei, Dario and Sutskever, Ilya},
  howpublished={\url{https://openai.com/research/language-unsupervised}},
  year={2019},
  note={OpenAI Blog}
}

@ARTICLE{11112781,
  author={Habib, Md Arafat and Iturria-Rivera, Pedro Enrique and Ozcan, Yigit and Elsayed, Medhat and Bavand, Majid and Gaigalas, Raimundas and Erol-Kantarci, Melike},
  journal={IEEE TNSE}, 
  title={Harnessing the Power of LLMs, Informers and Decision Transformers for Intent-Driven RAN Management in 6G}, 
  year={2025},
  volume={},
  number={},
  pages={1-20},
  doi={10.1109/TNSE.2025.3596028}}

@INPROCEEDINGS{9789886,
  author={Schneider, Stefan and Werner, Stefan and Khalili, Ramin and Hecker, Artur and Karl, Holger},
  booktitle={NOMS 2022-2022 IEEE/IFIP}, 
  title={mobile-env: An Open Platform for Reinforcement Learning in Wireless Mobile Networks}, 
  year={2022},
  volume={},
  number={},
  pages={1-3},
  doi={10.1109/NOMS54207.2022.9789886}}

@article{schneider2023deepcomp,
	title={Multi-Agent Deep Reinforcement Learning for Coordinated Multipoint in Mobile Networks},
	author={Schneider, Stefan and Karl, Holger and Khalili, Ramin and Hecker, Artur},
	journal={IEEE TNSM},
	year={2023},
}

@inproceedings{NEURIPS2020_1457c0d6,
 author = {Brown, Tom and Mann, Benjamin and Ryder, Nick and others},
 booktitle = {NeurIPS},
 pages = {1877--1901},
 title = {Language Models are Few-Shot Learners},
 volume = {33},
 year = {2020}
}

@mastersthesis{ali2024explainability,
  author       = {Hamza Ali},
  title        = {Explainability of Transformer-Based Reinforcement Learning},
  school       = {KTH Royal Institute of Technology},
  address      = {Stockholm, Sweden},
  year         = {2024},
  type         = {Master’s thesis},
  note         = {TRITA-EECS-EX; 2024:824},
  url          = {https://kth.diva-portal.org/smash/record.jsf?pid=diva2:1940023},
  urn          = {urn:nbn:se:kth:diva-360332}
}

@INPROCEEDINGS{10901312,
  author={Zhang, Han and Bin Sediq, Akram and Afana, Ali and Erol-Kantarci, Melike},
  booktitle={2024 IEEE GLOBECOM}, 
  title={Large Language Models in Wireless Application Design: In-Context Learning-enhanced Automatic Network Intrusion Detection}, 
  year={2024},
  volume={},
  number={},
  pages={2479-2484},
  doi={10.1109/GLOBECOM52923.2024.10901312}}

@INPROCEEDINGS{10901084,
  author={Xue, Nan and Sun, Yaping and Chen, Zhiyong and Tao, Meixia and Xu, Xiaodong and Qian, Liang and Cui, Shuguang and Zhang, Ping},
  booktitle={2024 IEEE GLOBECOM}, 
  title={WDMoE: Wireless Distributed Large Language Models with Mixture of Experts}, 
  year={2024},
  volume={},
  number={},
  pages={2707-2712},
  doi={10.1109/GLOBECOM52923.2024.10901084}}

@Article{app16041823,
AUTHOR = {Mu, Xian and Liu, Mingzhu and Xu, Yao and Li, Dagang},
TITLE = {CoMEx: Continual Mixture of Experts for Fast Policy Adaptation in RAN Slicing},
JOURNAL = {Applied Sciences},
VOLUME = {16},
YEAR = {2026},
NUMBER = {4},
ARTICLE-NUMBER = {1823},
URL = {https://www.mdpi.com/2076-3417/16/4/1823},
ISSN = {2076-3417},
DOI = {10.3390/app16041823}
}

@INPROCEEDINGS{10592370,
  author={Du, Hongyang and Liu, Guangyuan and Lin, Yijing and Niyato, Dusit and Kang, Jiawen and Xiong, Zehui and Kim, Dong In},
  booktitle={2024 IWCMC}, 
  title={Mixture of Experts for Intelligent Networks: A Large Language Model-enabled Approach}, 
  year={2024},
  volume={},
  number={},
  pages={531-536},
  doi={10.1109/IWCMC61514.2024.10592370}}

@article{xu2026moesurvey,
  AUTHOR = {Xu, Yunting and Wang, Jiacheng and Zhang, Ruichen and Zhao, Changyuan and Niyato, Dusit and Kang, Jiawen and Xiong, Zehui and Qian, Bo and Zhou, Haibo and Mao, Shiwen and Jamalipour, Abbas and Shen, Xuemin and Kim, Dong In},
  TITLE = {Mixture of Experts for Decentralized Generative AI and Reinforcement Learning in Wireless Networks: A Comprehensive Survey},
  JOURNAL = {IEEE COMST},
  VOLUME = {28},
  PAGES = {4051--4085},
  YEAR = {2026}
}

@ARTICLE{7839266,
  author={Bassoy, Selcuk and Farooq, Hasan and Imran, Muhammad A. and Imran, Ali},
  journal={IEEE COMST}, 
  title={Coordinated Multi-Point Clustering Schemes: A Survey}, 
  year={2017},
  volume={19},
  number={2},
  pages={743-764},
  doi={10.1109/COMST.2017.2662212}}

@techreport{nvidia2026nemotron3super,
  title       = {Nemotron 3 Super: Open, Efficient Mixture-of-Experts Hybrid Mamba-Transformer Model for Agentic Reasoning},
  author      = {{NVIDIA}},
  institution = {NVIDIA},
  year        = {2026},
  month       = apr,
  type        = {Technical Report},
  url         = {https://arxiv.org/pdf/2604.12374},
  urldate     = {2026-07-10},
  note        = {Also available as arXiv:2604.12374}
}

@ARTICLE{11240212,
  author={Liu, Zhaoyang and Wang, Xijun and Feng, Chenyuan and Sun, Xinghua and Zhan, Wen and Chen, Xiang},
  journal={IEEE TCOM}, 
  title={Meta-Reinforcement Learning With Mixture of Experts for Generalizable Multi Access in Heterogeneous Wireless Networks}, 
  year={2026},
  volume={74},
  number={},
  pages={870-885},
  doi={10.1109/TCOMM.2025.3631558}}

@ARTICLE{9427543,
  author={Solaija, Muhammad Sohaib J. and Salman, Hanadi and Kihero, Abuu B. and Sağlam, Mehmet Izzet and Arslan, Hüseyin},
  journal={IEEE Access}, 
  title={Generalized Coordinated Multipoint Framework for 5G and Beyond}, 
  year={2021},
  volume={9},
  number={},
  pages={72499-72515},
  doi={10.1109/ACCESS.2021.3079190}}

@ARTICLE{9681835,
  author={Elhattab, Mohamed and Arfaoui, Mohamed Amine and Assi, Chadi and Ghrayeb, Ali},
  journal={IEEE JSAC}, 
  title={RIS-Assisted Joint Transmission in a Two-Cell Downlink NOMA Cellular System}, 
  year={2022},
  volume={40},
  number={4},
  pages={1270-1286},
  doi={10.1109/JSAC.2022.3143211}}

@article{article_heuristic,
author = {Beylerian, Anthony and Ohtsuki, Tomoaki},
year = {2016},
month = {01},
pages = {},
title = {Multi-point fairness in resource allocation for C-RAN downlink CoMP transmission},
volume = {2016},
journal = {EURASIP Journal on Wireless Communications and Networking},
doi = {10.1186/s13638-015-0501-4}
}

@ARTICLE{11370176,
  author={Jiang, Feibo and Pan, Cunhua and Wang, Kezhi and Michiardi, Pietro and Dobre, Octavia A. and Debbah, Merouane},
  journal={IEEE JSAC}, 
  title={From Large AI Models to Agentic AI: A Tutorial on Future Intelligent Communications}, 
  year={2026},
  volume={},
  number={},
  pages={1-1},
  doi={10.1109/JSAC.2026.3660010}}

@inproceedings{
kong2025mastering,
title={Mastering Massive Multi-Task Reinforcement Learning via Mixture-of-Expert Decision Transformer},
author={Yilun Kong and Guozheng Ma and Qi Zhao and Haoyu Wang and Li Shen and Xueqian Wang and Dacheng Tao},
booktitle={ICML},
year={2025},
url={https://openreview.net/forum?id=qUcUyqP1UA}
}

@inproceedings{
wu2025mixtureofexperts,
title={Mixture-of-Experts Meets In-Context Reinforcement Learning},
author={Wenhao Wu and Fuhong Liu and Haoru Li and Zican Hu and Daoyi Dong and Chunlin Chen and Zhi Wang},
booktitle={NeurIPS},
year={2025},
url={https://openreview.net/forum?id=VMqxRPqdPw}
}

@InProceedings{harmoDT,
  title = {HarmoDT: Harmony Multi-Task Decision Transformer for Offline Reinforcement Learning},
  author = {Hu, Shengchao and Fan, Ziqing and Shen, Li and Zhang, Ya and Wang, Yanfeng and Tao, Dacheng},
  booktitle = {ICML},
  pages = {19182--19197},
  year = {2024},
  volume = {235},
  series = {PMLR},
  month = jul,
  publisher = {PMLR},
  url = {https://proceedings.mlr.press/v235/hu24d.html}
}

@article{schulman2017ppo,
  title={Proximal Policy Optimization Algorithms},
  author={Schulman, John and Wolski, Filip and Dhariwal, Prafulla and Radford, Alec and Klimov, Oleg},
  journal={arXiv preprint arXiv:1707.06347},
  year={2017}
}

@inproceedings{
bhargava2024when,
title={When should we prefer Decision Transformers for Offline Reinforcement Learning?},
author={Prajjwal Bhargava and Rohan Chitnis and Alborz Geramifard and Shagun Sodhani and Amy Zhang},
booktitle={ICLR},
year={2024},
url={https://openreview.net/forum?id=vpV7fOFQy4}
}

@inproceedings{Temiz2026Generalizable,
  author    = {Fatih Temiz and Shavbo Salehi and Melike Erol-Kantarci},
  title     = {Generalizable Multi-Task Learning for Wireless Networks Using Prompt Decision Transformers},
  booktitle = {Proceedings of the 2026 IEEE MeditCom},
  year      = {2026},
  month     = jul,
  address   = {Cagliari, Italy},
  publisher = {IEEE}
}

@inproceedings{Temiz2026EdgeLearning,
  author    = {F. Temiz and S. Salehi and M. Erol-Kantarci},
  title     = {Edge Learning via Federated Split Decision Transformers for Metaverse Resource Allocation},
  booktitle = {IEEE ICC},
  year      = {2026},
  organization = {IEEE},
}

@ARTICLE{10766614,
  author={Tsampazi, Maria and D'Oro, Salvatore and Polese, Michele and Bonati, Leonardo and Poitau, Gwenael and Healy, Michael and Alavirad, Mohammad and Melodia, Tommaso},
  journal={IEEE TMC}, 
  title={PandORA: Automated Design and Comprehensive Evaluation of Deep Reinforcement Learning Agents for Open RAN}, 
  year={2025},
  volume={24},
  number={4},
  pages={3223-3240},
  doi={10.1109/TMC.2024.3505781}}

@ARTICLE{10329947,
  author={Polese, Michele and Dohler, Mischa and Dressler, Falko and Erol-Kantarci, Melike and Jana, Rittwik and Knopp, Raymond and Melodia, Tommaso},
  journal={IEEE JSAC}, 
  title={Empowering the 6G Cellular Architecture With Open RAN}, 
  year={2024},
  volume={42},
  number={2},
  pages={245-262},
  doi={10.1109/JSAC.2023.3334610}}

@ARTICLE{10292755,
  author={Khan, Nasir and Coleri, Sinem and Abdallah, Asmaa and Celik, Abdulkadir and Eltawil, Ahmed M.},
  journal={IEEE Commun. Mag.}, 
  title={Explainable and Robust Artificial Intelligence for Trustworthy Resource Management in 6G Networks}, 
  year={2024},
  volume={62},
  number={4},
  pages={50-56},
  doi={10.1109/MCOM.001.2300172}}

@INPROCEEDINGS{851585,
  author={Medeisis, A. and Kajackas, A.},
  booktitle={VTC2000-Spring. 2000 IEEE 51st VTC}, 
  title={On the use of the universal Okumura-Hata propagation prediction model in rural areas}, 
  year={2000},
  volume={3},
  number={},
  pages={1815-1818 vol.3},
  doi={10.1109/VETECS.2000.851585}}

@ARTICLE{9737471,
  author={Elhattab, Mohamed and Arfaoui, Mohamed Amine and Assi, Chadi},
  journal={IEEE TCOM}, 
  title={Joint Clustering and Power Allocation in Coordinated Multipoint Assisted C-NOMA Cellular Networks}, 
  year={2022},
  volume={70},
  number={5},
  pages={3483-3498},
  doi={10.1109/TCOMM.2022.3160547}}

@ARTICLE{11535007,
  author={Han, Cao and Chen, Ran and Yao, Shuaizhen and Xu, Chunyan and Mei, Xiaowei},
  journal={IEEE OJVT}, 
  title={Frequency-Aware Decision Transformer for Offline Policy Learning in Adaptive Twin Control}, 
  year={2026},
  volume={7},
  number={},
  pages={1656-1666},
  doi={10.1109/OJVT.2026.3696721}}

@article{bengio2013estimating,
  title={Estimating or propagating gradients through stochastic neurons for conditional computation},
  author={Bengio, Yoshua and L{\'e}onard, Nicholas and Courville, Aaron},
  journal={arXiv preprint arXiv:1308.3432},
  year={2013}
}

@article{jiang2024mixtral,
  title={Mixtral of Experts},
  author={Jiang, Albert Q. and Sablayrolles, Alexandre and Roux, Antoine
          and Mensch, Arthur and Savary, Blanche and Bamford, Chris
          and Chaplot, Devendra Singh and de las Casas, Diego
          and Bou Hanna, Emma and Bressand, Florian and others},
  journal={arXiv preprint arXiv:2401.04088},
  year={2024}
}

@misc{yang2026freetokenefficientedgenativemoe,
      title={FreeToken: Efficient Edge-Native MoE Serving with Bandwidth-Adaptive Execution}, 
      author={Shuo Yang and Xiaoze Fan and Melissa Pan and Haocheng Xi and Zhe Wang and Shanlin Sun and Kurt Keutzer and Song Han and Matei Zaharia and Chenfeng Xu and Ion Stoica},
      year={2026},
      eprint={2608.16157},
      archivePrefix={arXiv},
      primaryClass={cs.DC},
      url={https://arxiv.org/abs/2608.16157}, 
}

@ARTICLE{11303878,
  author={Liu, Guangyuan and Du, Hongyang and Niyato, Dusit and Kang, Jiawen and Xiong, Zehui and Jamalipour, Abbas and Mao, Shiwen and Kim, Dong In},
  journal={IEEE Network}, 
  title={Fusion of Mixture of Experts and Generative Artificial Intelligence in Mobile Edge Metaverse}, 
  year={2025},
  volume={},
  number={},
  pages={1-9},
  doi={10.1109/MNET.2025.3639257}}

@ARTICLE{11417148,
  author={Ma, Jitong and Li, Jianing and Zhao, Sen and Li, Hui Dong and Dai, Jun Yan and Wang, Jie and Cheng, Qiang},
  journal={IEEE TCCN}, 
  title={Personalized Federated Learning With Mixture-of-Experts for Automatic Modulation Classification Under Data Heterogeneity}, 
  year={2026},
  volume={12},
  number={},
  pages={6478-6491},
  doi={10.1109/TCCN.2026.3668838}}

@ARTICLE{11527015,
  author={Liu, Tao and Xie, Qipeng and Han, Zhaoyang and Wang, Weizheng and Pan, Shengli and Zhang, Xiang and Su, Chunhua and Wu, Kaishun},
  journal={IEEE TCCN}, 
  title={Efficient Edge Deployment of Mixture-of-Experts Models With Hybrid Expert Quantization}, 
  year={2026},
  volume={12},
  number={},
  pages={10164-10178},
  doi={10.1109/TCCN.2026.3694872}}

\end{document}